\documentclass[letterpaper,twocolumn,english,prb,superscriptaddress,notitlepage]{revtex4-1}
\usepackage[latin9]{inputenc}
\usepackage{array}
\usepackage{multirow}
\usepackage{amsmath}
\usepackage{graphicx}

\makeatletter

\pdfpageheight\paperheight
\pdfpagewidth\paperwidth

\providecommand{\tabularnewline}{\\}

\usepackage{babel}
\usepackage{multirow}

\usepackage{xcolor}

\def\br{{\bf r}}
\def\bp{{\bf r}^\prime}

\makeatother

\usepackage{babel}
\begin{document}

\title{Simple eigenvalue-self-consistent $\bar{\Delta}GW_{0}$}

\author{Vojt\v{e}ch Vl\v{c}ek}
\email{vojtech.vlcek@gmail.com}

\affiliation{Department of Chemistry and Biochemistry, University of California,
Los Angeles California 90095, U.S.A.}

\affiliation{After July 1 2018: Department of Chemistry and Biochemistry, University
of California, Santa Barbara California 93106, U.S.A.}

\author{Roi Baer}
\email{roi.baer@huji.ac.il}

\affiliation{Fritz Haber Center for Molecular Dynamics, Institute of Chemistry,
The Hebrew University of Jerusalem, Jerusalem 91904, Israel}

\author{Eran Rabani}
\email{eran.rabani@berkeley.edu}

\affiliation{Department of Chemistry, University of California and Materials Science
Division, Lawrence Berkeley National Laboratory, Berkeley, California
94720, USA}

\affiliation{The Raymond and Beverly Sackler Center for Computational Molecular
and Materials Science, Tel Aviv University, Tel Aviv, Israel 69978}

\author{Daniel Neuhauser}
\email{dxn@ucla.edu}

\affiliation{Department of Chemistry and Biochemistry, University of California,
Los Angeles California 90095, U.S.A.}
\begin{abstract}
We derive a general form of eigenvalue self-consistency for $GW_{0}$
in the time domain and use it to obtain a simplified postprocessing
eigenvalue self-consistency, which we label $\bar{\Delta}GW_{0}$.  The
method costs the same as a one-shot $G_{0}W_{0}$ when the latter gives
the full frequency-domain (or time-domain) matrix element of the
self-energy. The accuracy of $\bar{\Delta}GW_{0}$ increases with
system size, as demonstrated here by comparison to other $GW$
self-consistency results and to CCSD(T) predictions. When combined
with the large-scale stochastic $G_{0}W_{0}$ formulation
$\bar{\Delta}GW_{0}$ is applicable to very large systems, as
exemplified by periodic supercells of semiconductors and insulators
with 2048 valence electrons. For molecules the error of our eventual
partially self-consistent approach starts at about 0.2eV for small
molecules and decreases to 0.05eV for large ones, while for the
periodic solids studied here the mean-absolute-error is only 0.03eV.
\end{abstract}
\maketitle

\section{Introduction}

The $GW$ approximation~\cite{Hedin1965} to many-body perturbation
theory is often used to calculate electron removal or addition
energies and related (inverse) photoemission spectra of molecules,
nanostructures, and bulk
materials.\cite{Aryasetiawan1998,Rieger1999,Steinbeck1999,Onida2002,Rubio2005,Friedrich2006,Shishkin2007,Trevisanutto2008,Rostgaard2010,Tamblyn2011,Setten2012,Stefanucci2013,Govoni2015,Kaplan2016}
$GW$ is part of a family of methods that describe the probability
amplitude of a quasiparticle (QP) to propagate between two space-time
points $\left(\br,t\right)$ and $\left(\bp,t'\right)$ with a Green's
function $G\left(\br,\bp,t,t'\right)$, the poles of which are the QP
energies. The Green's function is obtained perturbatively from a
reference (non-interacting) Green's function,
$G_{0}\left(\br,\bp,t,t'\right)$, via a Dyson equation (all equations
use atomic units):
\begin{align}
 & {G}\left(\br,\bp,t,t'\right)=G_{0}\left(\br,\bp,t,t'\right)+\iint
  G_{0}\left(\br,\br_{1},t,t_{1}\right)\times\nonumber \\ &
  \Sigma\left(\br_{1},\br_{2},t_{1}-t_{2}\right)G\left(\br_{2},\bp,t_{2},t'\right)d\br_{1}dt_{1}d\br_{2}dt_{2},\label{eq:dyson}
\end{align}
where $\Sigma\left(\br_{1},\br_{2},t,t'\right)$ represents the self-energy.
The reference Green's function is typically\cite{Hybertsen1985,Hybertsen1986}
given by the Kohn-Sham~\cite{Kohn1965} (KS) density function theory
(DFT).\cite{Hohenberg1964} In $GW$ , the self-energy $\Sigma$ is
approximated as: 
\begin{equation}
\Sigma\left(\br,\bp,t\right)=iG\left(\br,\bp,t\right)W\left(\br,\bp,t^{+}\right),\label{eq:gw}
\end{equation}
where $W\left(\br,\bp,t\right)$ is the screened Coulomb interaction,
usually evaluated within the random phase approximation (RPA).

A solution of the equations above in requires, in principle, a
self-consistent procedure since both $\Sigma\left(\br,\bp,t\right)$
and $W\left(\br,\bp,t^{+}\right)$ depend on
$G\left(\br,\bp,t\right)$. In practice, the self-consistency is often
abandoned and the most common $GW$ treatment is based on ``one-shot''
scheme\cite{Hybertsen1985,Hybertsen1986}, which we label as
$G_{0}W_{0}$, since the right hand side of Eq.~\eqref{eq:gw} becomes
$G_{0}\left(\br,\bp,t\right)W_{0}\left(\br,\bp,t\right)$, where
$W_{0}$ is obtained from a random phase approximation that uses the KS
eigenstates. The one-shot approach improves significantly the KS-DFT
results, yet it depends on the choice of the reference system and
often underestimates the QP gaps ($E_{g}$) and the ionization
potentials
($I$).\cite{Faleev2004,Shishkin2007,Caruso2012,BrunevalMarques2012,Bruneval2012,Setten2015,Knight2016}
A fully self-consistent solution is computationally extremely
demanding\cite{stan2006fully,StanVanLeeuwen2009,RostgaardJacobsenThygesen2010,Blase2011,Deslippe2012,Nguyen2012,caruso2012unified,caruso2013self,koval2014fully,wang2015fully}
and in many situations it yields results that are worse than
$G_{0}W_{0}$.\cite{holm1998fully,Shishkin2007,BrunevalGatti2014}

To simplify the problem, a static and Hermitian approximation to the
self-energy \cite{Faleev2004,van2006quasiparticle} is sometimes used
in the so-called QP self-consistent $GW$ (qp$GW$), which iteratively
updates $\Sigma$ and QP wave-functions (i.e., Dyson orbitals). The
qp$GW$ approach is still computationally expensive and cannot be
applied for large systems; further it tends to overestimate $E_{g}$
\cite{van2006quasiparticle,kotani2007quasiparticle,BrunevalGatti2014}
and the ionization potentials
\cite{kaplan2015off,kaplan2016quasi,caruso2016benchmark}, due to an
overly strong screened Coulomb
interaction. \cite{van2006quasiparticle,BrunevalGatti2014,kaplan2016quasi}
Alternatively, the $W$ term is kept ``frozen'' and self-consistency is
sought only in the Green's
function.\cite{Shishkin2007,shishkin2007self,StanVanLeeuwen2009,Blase2011,BrunevalGatti2014,kaplan2015off}
This method is termed eigenvalue self-consistent $GW$ (ev$GW_{0}$) and
it was applied successfully to bulk
systems\cite{Northrup1987,Shishkin2007,BrunevalGatti2014} and to
organic molecules,\cite{StanVanLeeuwen2009,Blase2011,Knight2016} with
remarkable success. Even though it is cheaper than other
self-consistency methods, $\Sigma$ has to be recalculated in each
iteration, making ev$GW_{0}$ out of reach for nanoscale systems with
thousands of occupied electronic states.

Here a time domain formulation (Sec.~\ref{sec:Theory}) is used to
derive a simplified ev$GW_{0}$ formalism, labeled
$\bar{\Delta}GW_{0}$, where the self-consistency is only a
postprocessing step. Hence, as long as one has access to the matrix
element of the self-energy at all frequencies or all times, then,
irrespective of system size, the computational cost of the
self-consistency is negligible (i.e., seconds on a single-core
machine) so $\bar{\Delta}GW_{0}$ costs not more than $G_{0}W_{0}.$

We specifically combine $\bar{\Delta}GW$ with our stochastic
$G_{0}W_{0}$ approach, which has a nearly linear
scaling\cite{Neuhauser2014,vlcek2017stochastic} and enables
$G_{0}W_{0}$ for extremely large
systems\cite{Vlcek2016,vlcek2018quasiparticle}.  The stochastic
$G_{0}W_{0}$ method has automatically the necessary ingredient for
$\bar{\Delta}GW_{0}$, as it produces the matrix element of the
self-energy at all times.

The combined method (stochastic $G_{0}W_{0}$ with
$\bar{\Delta}GW_{0}$) is first tested in
Sec.~\ref{sec:Results-and-Discussion} on molecules, and we find that
$\bar{\Delta}GW$ becomes more accurate as the system size
increases. Next, we perform stochastic $G_{0}W_{0}$ calculations for
periodic semiconductors and insulators using large supercells with
2048 valence electrons. For solids $\bar{\Delta}GW_{0}$ gives $E_{g}$
in excellent agreement with experiment and a mean absolute error of
0.03~eV.

In all cases the self-consistency is reached in very few iterations
without any additional cost on top of the $G_{0}W_{0}$ step.

\section{\label{sec:Theory}Theory}

\subsection{Green's function self-consistency in the time domain}

The QP energy of the $i^{{\rm th}}$ state
$\varepsilon^{QP}\left(i\right)$ is calculated using the usual form of
the perturbative $GW$ approximation in which the Kohn-Sham eigenvalues
($\varepsilon^{0}$) are corrected by the QP shift ($\Delta$) using a
fixed point equation:
\begin{equation}
\varepsilon^{QP}\left(i\right)=\varepsilon^{0}\left(i\right)+\Delta\left(i\right),\label{QPEq}
\end{equation}
where 
\begin{equation}
\Delta\left(i\right)=\tilde{\Sigma}_{i}\left[\omega=\varepsilon^{QP}\left(i\right)\right]-\left\langle
\phi_{i}\left|v_{xc}\right|\phi_{i}\right\rangle .\label{QPShift}
\end{equation}

Here, $v_{xc}$ is the Kohn-Sham exchange-correlation potential for the
DFT density, $\tilde{\Sigma}_{i}\left[\omega\right]$ is the Fourier
transform of the matrix-element of $\hat{\Sigma}(t)$
\begin{equation}
\tilde{\Sigma}_{i}\left[\omega\right]=\int\left\langle
\phi_{i}\middle|\hat{\Sigma}\left(t\right)\middle|\phi_{i}\right\rangle
e^{i\omega t}dt,
\end{equation}
and $\hat{\Sigma}(t)$ is given by Eq.~\eqref{eq:gw}.

Starting from a KS DFT reference point, the initial self-energy is
constructed from the KS propagator
\begin{align}
iG_{0}\left(\br,\br',t\right) & ={\rm Tr}\left\{
\left|\br\right\rangle \left\langle
\bp\right|e^{-i\hat{h}^{0}t}\right.\nonumber \\ &
\left.\left[\theta\left(t\right)\theta_{\beta}\left(\hat{h}^{0}-\mu\right)-\theta\left(-t\right)\theta_{\beta}\left(\mu-\hat{h}^{0}\right)\right]\right\}
,
\end{align}
where ${\rm Tr}$ denotes a trace over all KS states, $\mu$ is the
chemical potential, $\theta$ is the Heaviside step function that
guarantees forward and backward time propagation for particles and
holes, respectively, and $\hat{h}^{0}$ is the KS Hamiltonian
\begin{equation}
\hat{h}^{0}=-\frac{1}{2}\nabla^{2}+v_{{\rm ext}}+v_{{\rm H}}+v_{\text{xc}},
\end{equation}
where we introduced the kinetic energy and the external and Hartree
potentials. In the rest of the paper we employ real time-dependent
Hartree propagation to calculate the screened Coulomb interaction
\cite{Neuhauser2014,vlcek2017stochastic}; this is equivalent to using
the RPA approximation for $W_{0}$.

In the time-domain, the self-energy matrix element for the $i^{{\rm th}}$
state is 
\begin{align}
 & \left\langle
  \phi_{i}\middle|\hat{\Sigma}_{0}\left(t\right)\middle|\phi_{i}\right\rangle
  =\\ &
  i\iint\phi_{i}\left(\br\right)G_{0}\left(\br,\bp,t\right)W_{0}\left(\br,\bp,t^{+}\right)\phi_{i}\left(\bp\right)d\br
  d\bp.\nonumber
\end{align}
Finally, after Fourier transformation combined with time-ordering
\cite{Neuhauser2014,vlcek2017stochastic} the ``one-shot'' QP energy
is calculated through Eq.~\eqref{QPShift}.

In the ev$GW_{0}$ procedure, the Green's function is reconstructed in
each iteration, employing the QP energies from the previous iteration.
In the time domain this corresponds to writing the propagator as
\begin{align}
iG\left(\br,\br',t\right) & ={\rm Tr}\left\{ \left|\br\right\rangle
\left\langle
\bp\right|e^{-i\left(\hat{h}^{0}+\hat{\Delta}\right)t}\right.\nonumber
\\ &
\left.\left[\theta\left(t\right)\theta_{\beta}\left(\hat{h}^{0}-\mu\right)-\theta\left(-t\right)\theta_{\beta}\left(\mu-\hat{h}^{0}\right)\right]\right\}
,\label{eq:GwithDelta}
\end{align}
where $\hat{\Delta}$ contains all the many-body contributions.

As common in ev$GW_{0}$ self-consistency
\cite{shishkin2007self,BrunevalGatti2014,kaplan2015off}, the fact that
the true self-energy operator is non-Hermitian and non-diagonal is
disregarded and, for the purposes of Eq.~\eqref{eq:GwithDelta}, it is
expressed in the KS basis as
\begin{equation}
\hat{\Delta}\simeq\sum_{i}\left|\phi_{i}\middle\rangle{\rm Re}\Delta\left(i\right)\middle\langle\phi_{i}\right|.\label{eq:Deltasum}
\end{equation}
Hence, all the KS energies in the exponent in
Eq.~\eqref{eq:GwithDelta} are shifted to the QP energies obtained from
Eq.~\eqref{QPEq}. Eq.~\eqref{eq:Deltasum} is basically the fundamental
eqaution of ev$GW_{0}$.

We now note that the operator $\hat{\Delta}$ in
Eq.~\eqref{eq:GwithDelta} is by construction diagonal in the KS basis
set and we thus express it as a function of the KS Hamiltonian
$\bar{\Delta}\left(\hat{h}^{0}\right)$ that interpolates all QP
shifts. The Green's function is therefore:
\begin{align}
iG\left(\br,\br',t\right) & ={\rm Tr}\left\{ \left|\br\right\rangle
\left\langle
\bp\right|e^{-i\left[\hat{h}^{0}+\bar{\Delta}\left(\hat{h}^{0}\right)\right]t}\right.\nonumber
\\ &
\left.\left[\theta\left(t\right)\theta_{\beta}\left(\hat{h}^{0}-\mu\right)-\theta\left(-t\right)\theta_{\beta}\left(\mu-\hat{h}^{0}\right)\right]\right\}
.\label{eq:Gt}
\end{align}
This simple expression allows for a further approximation described
below that significantly reduces the computational cost associated
with self-consistent treatment.

\subsection{Efficient and inexpensive implementation}

In many cases $\bar{\Delta}$ is well described by a low degree
polynomial with discontinuity at the band gap energies,
$\varepsilon\left(H\right)$ and $\varepsilon\left(L\right),$
corresponding to the highest occupied ($H$) and lowest unoccupied
($L$) states, respectively. The zeroth order term in this polynomial
corresponds to a scissors operator\cite{Filip2014,Qian2015} which
shifts the occupied and unoccupied states down and up in energy,
respectively
\begin{equation}
{\rm
  Re}\bar{\Delta\!}\left[\varepsilon^{0}\left(i\right)\right]\approx\begin{cases}
\Delta\!\left(H\right) &
\varepsilon^{0}\left(i\right)\le\varepsilon^{0}\!\left(H\right)\\ \Delta\!\left(L\right)
& \varepsilon^{0}\left(i\right)\ge\varepsilon^{0}\!\left(L\right)
\end{cases}.\label{scissors}
\end{equation}
We call this approximation $\bar{\Delta}GW_{0}$ and use it in
Sec.~\ref{sec:Results-and-Discussion} for molecules and periodic
systems.

\begin{figure}
\includegraphics[width=3.33in]{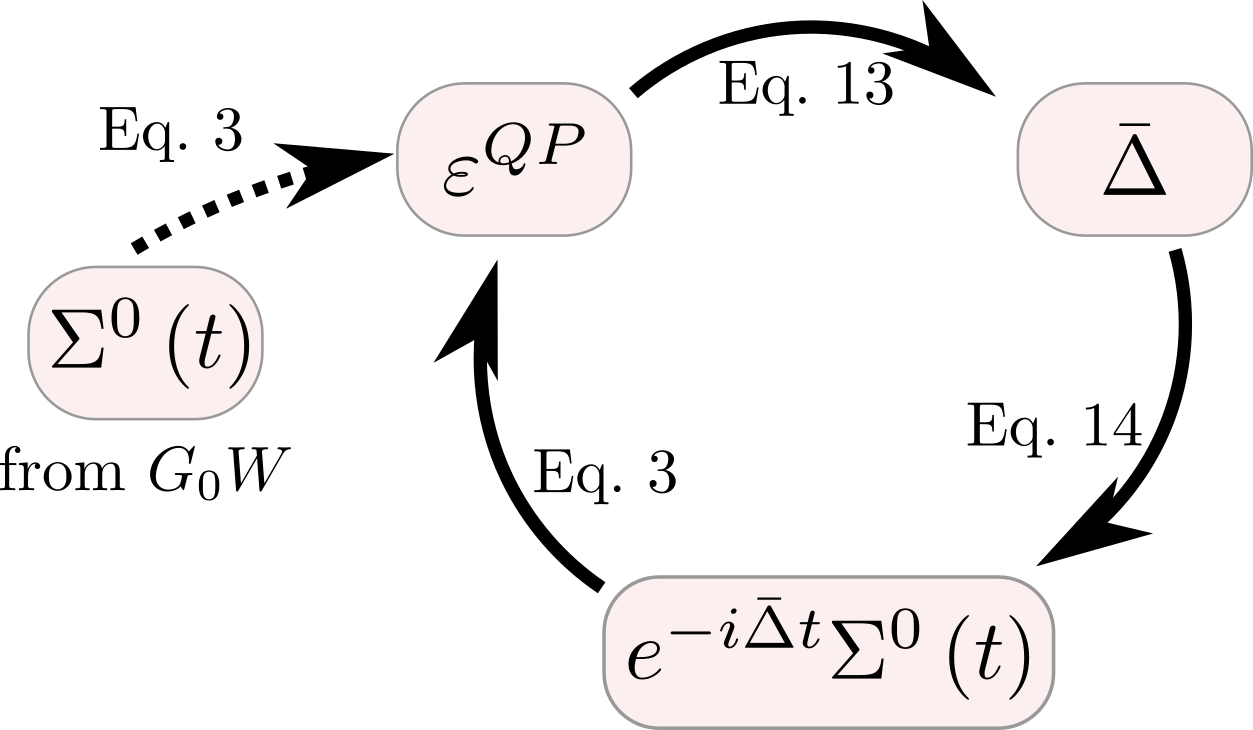} \caption{Illustration of
  the self-consistency cycle (full arrows). In the first step
  $G_{0}W_{0}$ is used to calculate the self-energy
  $\Sigma^{0}\left(t\right)$ and the corresponding QP energies
  $\varepsilon^{QP}$ for the HOMO and LUMO states (dashed arrow). The
  shift of occupied and unoccupied states ($\bar{\Delta}$) is
  calculated from the QP HOMO and LUMO energies through
  Eq.~\eqref{scissors}. The updated time-dependent self-energy
  $\Sigma\left(t\right)=e^{-i\bar{\Delta}t}\Sigma^{0}\left(t\right)$
  is obtained via Eq.~\eqref{tdscissors}, which leads to new QP
  HOMO/LUMO energies. The cycle is repeated a few times until reaching
  self-consistency. }
\label{flowchart} 
\end{figure}

Combining Eqs.~\eqref{scissors} and \eqref{eq:Gt} leads to a modified
Green's function which acquires an additional phase shift that is
different for positive and negative times. Therefore, in the time
domain we can define:
\begin{equation}
\bar{\Delta}\left(t\right)\equiv\begin{cases}
{\Delta}\!\left(H\right) & t<0\\
{\Delta}\!\left(L\right) & t>0
\end{cases}.\label{tdscissors}
\end{equation}
In each iteration, the updated self-energy matrix element is then
calculated as 
\begin{equation}
\left\langle
\phi_{i}\middle|\hat{\Sigma}\left(t\right)\middle|\phi_{i}\right\rangle
=e^{-i\bar{\Delta}\left(t\right)t}\left\langle
\phi_{i}\middle|\hat{\Sigma}_{0}\left(t\right)\middle|\phi_{i}\right\rangle
.
\end{equation}
Next, the self-energy matrix element is transformed to the frequency
domain and used in Eq.~\eqref{QPEq} to calculate a new estimate of the
QP energy. The new QP energy is used iteratively to update
Eqs.~\eqref{scissors} and \eqref{tdscissors}. The full cycle is
illustrated in Fig.~\ref{flowchart}.

Note that this form of self-consistency is trivial and is a
postprocessing step with \emph{no additional cost}, unlike previous
uses\cite{Filip2014,Qian2015} of the scissors-operator in $GW$ which
require repeated evaluations of the self-energy. Further, the
$\bar{\Delta}GW_{0}$ approach is applicable to any implementation
which yields $\Sigma\left(\omega\right)$.  It is naturally suited for
the stochastic $G_{0}W_{0}$ method
\cite{Neuhauser2014,vlcek2017stochastic} which provides the
self-energy on the full-time domain and therefore on a wide range of
frequencies (spanning several hundred eV).

\section{\label{sec:Results-and-Discussion}Results and Discussion}

\subsection{Molecules}

We first test our approach on ionization potentials $I$ (taken as
$-\varepsilon_{H}$) for a set of small molecules listed in
Table.~\ref{tab_molecules}.  A ground state DFT calculation is
performed using a Fourier real-space grid, ensuring (using the
Martyna-Tuckerman approach)\cite{martyna1999reciprocal} that the
potentials are not periodic. The exchange-correlation interaction is
described by local density approximation (LDA) \cite{PerdewWang} with
Troullier-Martins pseudopotentials\cite{TroullierMartins1991}; the DFT
eigenvalues are converged up to$<$10~meV with respect to the spacings
of the real space grids (given in Table~\ref{tab_molecules}).

\begin{table*}
\begin{tabular}{|c|c|c|cc|cc|c|c|c|c|}
\hline \multirow{2}{*}{system} & \multirow{2}{*}{$h$ ($a_{0}$)} &
\multicolumn{9}{c|}{$I({\rm eV)}$ }\tabularnewline \cline{3-11} & &
LDA & \multicolumn{2}{c|}{$G_{0}W$} &
\multicolumn{2}{c|}{$\bar{\Delta}GW_{0}$} & ev$GW_{0}$ & qp$GW$ &
CCSD(T) & Exp.\tabularnewline \hline \hline nitrogen & 0.35 & 10.44 &
15.08 & (0.05) & 15.93 & (0.05) &
15.32\footnote{Ref.~\onlinecite{caruso2016benchmark}} & 16.01$^{{\rm
    a}}$ & 15.57\footnote{Ref.~\onlinecite{krause2015coupled}} & 15.58
\tabularnewline \hline ethylene & 0.35 & 6.92 & 10.50 & (0.04) & 10.87
& (0.04) & 10.24$^{{\rm a}}$ & 10.63$^{{\rm a}}$ & 10.67$^{{\rm b}}$ &
10.68 \tabularnewline \hline urea & 0.30 & 6.10 & 9.53 & (0.08) &
10.48 & (0.08) & 9.81$^{{\rm a}}$ & 10.45$^{{\rm a}}$ & 10.05$^{{\rm
    b}}$ & 10.28 \tabularnewline \hline naphtalene & 0.35 & 5.71 &
8.10 & (0.09) & 8.39 & (0.09) &
8.15\footnote{Ref.~\onlinecite{rangel2016evaluating}} & - &
8.25$^{{\rm c}}$ & 8.14\tabularnewline \hline tetracene & 0.35 & 4.89
& 6.79 & (0.08) & 6.94 & (0.08) & 6.84$^{{\rm c}}$ & - & 7.02$^{{\rm
    c}}$ & 6.97\tabularnewline \hline hexacene & 0.35 & 4.52 & 6.15 &
(0.06) & 6.33 & (0.06) & 6.19$^{{\rm c}}$ & - & 6.32$^{{\rm c}}$ &
6.33\tabularnewline \hline
\end{tabular}

\caption{Ionization potentials $I$ for small molecules as calculated
  by different levels of theory. The $G_{0}W_{0}$ and
  $\bar{\Delta}GW_{0}$ estimates were obtained using an LDA starting
  point and calculated using a stochastic
  implementation\cite{Neuhauser2014,vlcek2017stochastic} with the
  statistical errors reported in parentheses. Experimental values are
  from Ref.~~\protect\onlinecite{NIST} . For several of the acenes the
  CCSD(T) results are estimates extrapolated to the infinite basis-set
  limit.\cite{rangel2016evaluating} Molecular geometries were taken
  from Refs.~\protect\onlinecite{NIST} and
  \protect\onlinecite{rangel2016evaluating}. }

\label{tab_molecules}
\end{table*}

The systems listed in the table are ordered according to the number
of valence electrons; N$_{2}$ and hexacene are the smallest and the
largest molecules studied here. In all cases, the stochastic $G_{0}W_{0}$
approach\cite{Neuhauser2014,vlcek2017stochastic} was used to calculate
the self-energy. We compare our calculations with reference values
taken from experiment and from CCSD(T). The geometries of the acene
molecules are taken from the $GW$ and CCSD(T) benchmark in Ref.~\onlinecite{rangel2016evaluating}.

Compared to the LDA eigenvalues, one-shot $G_{0}W$ predictions for the
ionization potentials are much closer to the CCSD(T) values with a
mean absolute error of 0.29~eV. In all cases, the value of $I$ is
underestimated, in agreement with previous benchmark
studies\cite{vansetten2015gw,vlcek2017stochastic,rangel2016evaluating}.
As the system size increases the difference between the $G_{0}W$ and
CCSD(T) values decreases so that the one-shot correction is an
increasingly better approximation.

\begin{figure}
\includegraphics[width=3.33in]{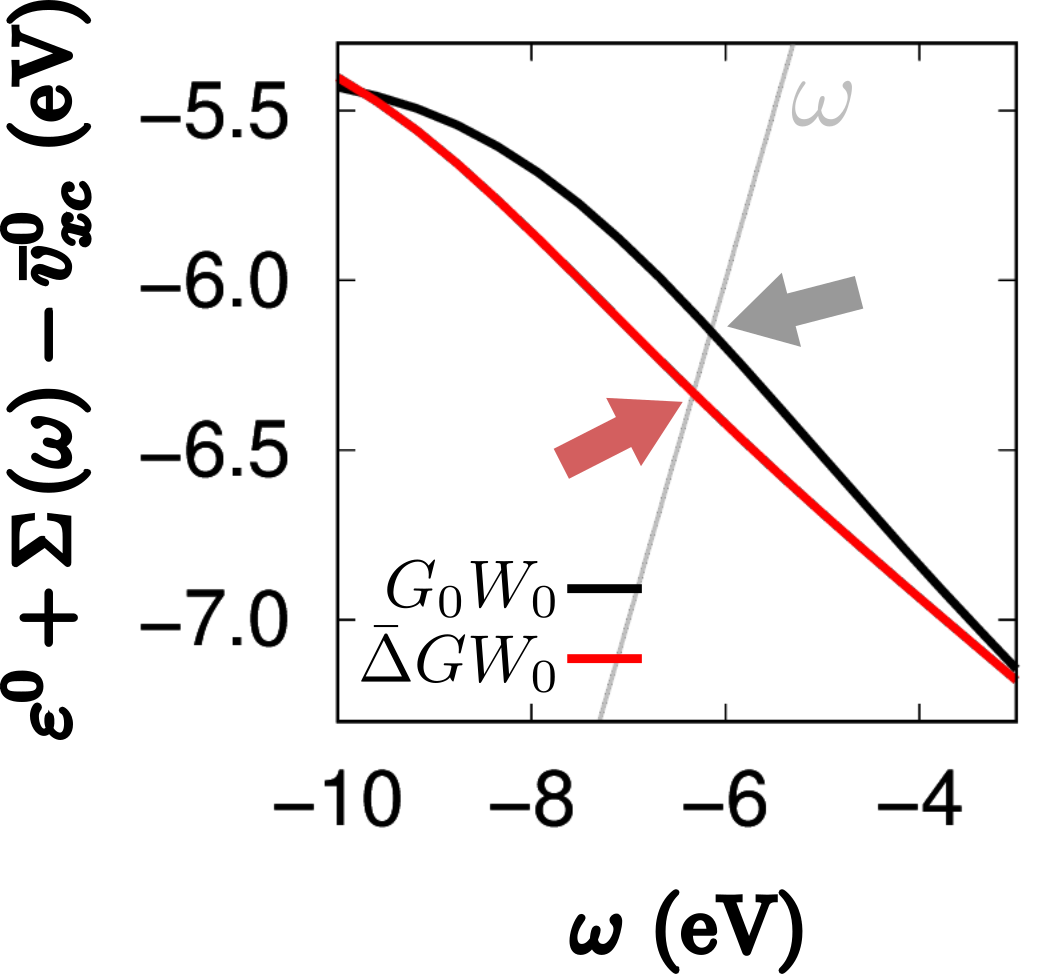} \caption{Graphical solution
  to the QP equation for hexacene for a self-energy from $G_{0}W_{0}$
  and $\bar{\Delta}GW_{0}$ . The gray line shows the frequency and the
  arrows indicate solutions to the fixed point equation
  (Eq.~\eqref{QPEq}) and give the QP energy.\textbf{ }The
  $\bar{\Delta}GW$ result is visibly shifted toward a lower energy
  than the one-shot solution. }
\label{sigma_hexac} 
\end{figure}

The simplified eigenvalue self-consistency converges in 3-4 iterations
after the initial $G_{0}W$ calculation; the initial and final
self-energy curves are illustrated for hexacene in
Fig.~\ref{sigma_hexac}. Since our self-consistency procedure is merely
a postprocessing step, its computational cost is negligible (less than
a second on single core machine).

We first compare our $\bar{\Delta}GW$ results with previous eigenvalue
and quasiparticle self-consistent $GW$ treatments (ev$GW$ and qp$GW$,
respectively). All methods consistently increase ionization potentials
above the one-shot values. The $\bar{\Delta}GW_{0}$ estimates are
higher than ev$GW$, but appear to be closer to qp$GW$. As there are
very few published qp$GW$ results for molecules, it is not possible to
assess whether this is a general trend for molecules.

For all the studied molecules our method yields results in good
agreement with CCSD(T). The improvement is only modest for the
smallest molecules, since the QP shift strongly depends on
$\varepsilon^{0}$, i.e., it is not constant for all occupied (or
unoccupied) states and the assumption of Eq.~\eqref{scissors} is not
fulfilled. For instance, for N$_{2}$ $G_{0}W_{0}$ shifts the lowest
valence state by -7.34~$\pm$0.06~eV, but the HOMO energy is decreased
by -4.63$\pm$0.05~eV. In contrast, the shifts are closer for hexacene:
$-2.73\pm0.08$ and $-1.63\pm0.06\,{\rm eV}$~ for the bottom valence
and HOMO states, respectively.

The mean absolute difference between the $\bar{\Delta}GW_{0}$ and
CCSD(T) values is 0.20~eV for molecules, but for the largest systems
(tetracene and hexacene) it is only $0.05$~eV. This indicates that (i)
the scissors operator approximation in Eq.~\eqref{scissors} is more
appropriate for larger molecules and (ii) $\bar{\Delta}GW_{0}$
self-consistency is more accurate when $G_{0}W_{0}$ is already a good
approximation, i.e., in our case it gives results that are
sufficiently close to the CCSD(T) values.

\subsection{Periodic systems}

Next we study self-consistency for several periodic solids listed
in Table~\ref{tab_solids} where we calculate the fundamental band
gaps 
\begin{equation}
E_{g}=\varepsilon_{L}-\varepsilon_{H}.
\end{equation}
The stochastic approach is extended here to treat periodic boundary
conditions\cite{Vlcek20182b}. We again employ LDA with
Troullier-Martins pseudopotentials and Fourier real-space grids with a
spacing $h$ which is sufficiently small that the eigenvalues are
converged to $<10$~meV (see Table~\ref{tab_solids}). The method is
demonstrated on large supercells with 512 atoms (corresponding to
$4\times4\times4$ conventional cells with 2048 valence electrons).

As mentioned earlier, the $G_{0}W_{0}$ treatment of such large systems
is enabled by the stochastic approach
\cite{Neuhauser2014,vlcek2017stochastic,Vlcek20182b}, but our
self-consistency scheme is applicable to any $G_{0}W_{0}$
implementation that yields the full-frequency or full-time matrix
element of the self-energy.

\begin{figure}
\includegraphics[width=3.33in]{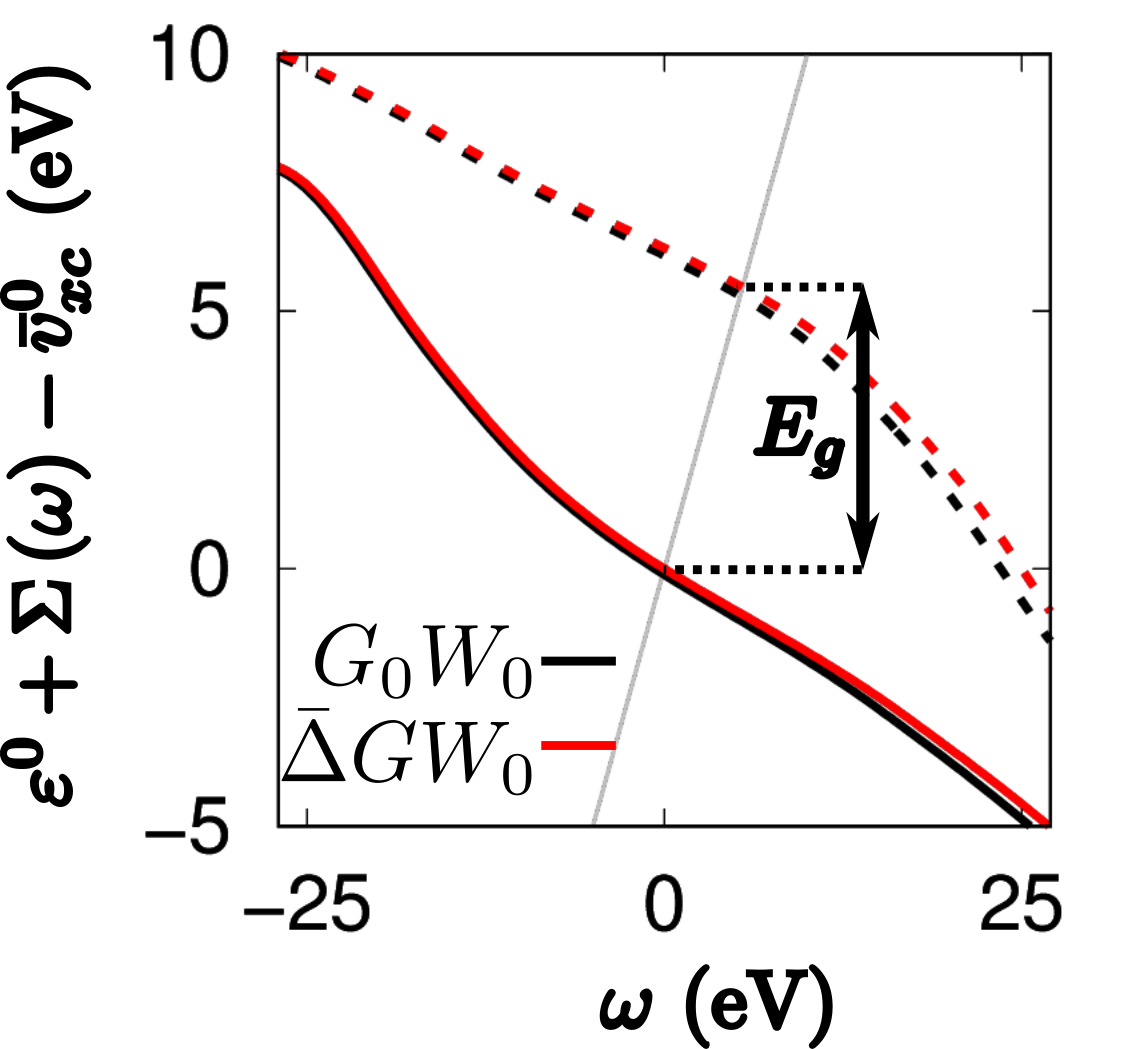} \caption{Graphical solution
  to the QP equation for the $G_{0}W_{0}$ and $\bar{\Delta}GW_{0}$
  self-energies, for a diamond solid simulated by a $4\times4\times4$
  supercell. The gray line denotes the frequency and the intersections
  with the black and red lines are the solution to the fixed point
  equation (Eq.~\eqref{QPEq}). The full and dashed lines are for the
  top valence and bottom conduction states, respectively.\textbf{
  }Both axes are shifted so that zero is associated with the QP energy
  of the top of the valence band as obtained from
  $\bar{\Delta}GW_{0}$. The self-consistent result is shifted with a
  larger band gap ($E_{g}$) than the one-shot solution.}
\label{sigma_diam} 
\end{figure}

The results in Table~\ref{tab_solids} show that the one-shot
correction yields band gaps that are lower than experimental values,
in agreement with previous
calculations. \cite{Faleev2004,van2006quasiparticle,Shishkin2007,shishkin2007self,BrunevalGatti2014}
In all cases studied, self-consistency is quickly achieved within 3 or
4 iterations. The resulting fundamental band gaps are enlarged by as
much as 0.20~eV,\textbf{ }and are quite close to experiment.  The
effect of using $\bar{\Delta}GW_{0}$ on the self-energy curves is
illustrated in Fig.~\ref{sigma_diam}.

Comparison to previous results (Table~\ref{tab_solids}) shows that the
$\bar{\Delta}GW_{0}$ fundamental gaps are overall at least as good as
the full eigenvalue self-consistency predictions. In contrast, qp$GW$
band gaps are too high and overestimate experiment by $\sim$10\% (also
see Ref.~\cite{VanHoucke2017}).

The $\bar{\Delta}GW_{0}$ results reproduce well the experimental
values, with the exception of bulk silicon. Note, however, that we
employ $4\times4\times4$ conventional cells with $\Gamma$ point
sampling. For silicon, this cell, while very large, is still not large
enough to reach the bulk limit. We can still compare our result with
the experimental $\Gamma-X$ gap which is higher (1.3~eV
\cite{tiago2004effect}).  Therefore, overall, for the set of solids
investigated the simplified self-consistency of $\bar{\Delta}GW_{0}$
yields gaps with excellent mean absolute error of 0.03~eV with respect
to experiment.

In all the investigated cases the difference between the QP shifts for
the bottom and top valence states are small and correlate slightly
with $E_{g}$. For BN we observe that in the zeroth iteration
($G_{0}W_{0}$) the bottom valence state and $\varepsilon_{H}$ are
shifted by $-0.73\pm0.04$ and $-0.21\pm0.03\,{\rm eV}$. However, for
Si the shifts are $0.13\pm0.03$ and $-0.16\pm0.02$~eV.

The excellent performance of $\bar{\Delta}GW$ is surprising but not
fortuitous. The structure of the self-energy curve is dominated by
plasmon
poles\cite{HybertsenLouie,godby1989metal,Aryasetiawan1998,larson2013role},
and the energy of these poles is proportional to the band gap. The
main goal of the iterative treatment is to capture the necessary
changes in the plasmon energy. We accomplish this goal by employing a
relative shift of occupied vs.~unoccupied states that acts like a
scissors operator (Eq.~\eqref{scissors}) that opens up the band gap
and leads to the desired increase in the plasmon frequency.

To test the dependence of the self-energy on energy and its
implications for the self-consistency, we further performed a set of
complementary calculations for two nanorystals, Si$_{35}$H$_{36}$ and
Si$_{705}$H$_{300}$, studied by stochastic $GW$ in the
past\cite{Neuhauser2014,vlcek2018quasiparticle}.  For these crystals
we did several calculations at different Kohn-Sham energies at fitted
the QP correction by a quadratic polynomial:
\begin{equation}
\bar{\Delta}\left(\varepsilon\right)=a_{2}\left(\varepsilon\right)^{2}+a_{1}\varepsilon+a_{0}.
\end{equation}
For the smaller system the QP correction terms are
$a_{2}=0.01$~eV$^{-1}$, $a_{1}=-0.22$ and $a_{0}=-1.13$~eV so
$\bar{\Delta}$ is far form being a constant. For the large Si
nanocrystal, which is already bulk\textendash like, we find however
that $a_{2}=0.00$~eV$^{-1}$, $a_{1}=0.04$ and $a_{0}=-1.52$~eV,
indicating only a weak linear dependence. We conclude that the
approximation of rigid shifts of all occupied and all unoccupied
states is well justified for solids, for which the variation of the QP
shift across the occupied states is much smaller than for molecular
systems, making the scissors-like assumption appropriate.

\begin{table*}
\begin{tabular}{|c|c|c|cc|cc|c|c|c|}
\hline \multirow{2}{*}{system} & \multirow{2}{*}{$h$ ($a_{0}$)} &
\multicolumn{8}{c|}{$E_{g}$ (eV)}\tabularnewline \cline{3-10} & & LDA
& \multicolumn{2}{c|}{$G_{0}W_{0}$} &
\multicolumn{2}{c|}{$\bar{\Delta}GW_{0}$} & ev$GW$
(Ref.~\onlinecite{shishkin2007self}) & qp$GW$
(Ref.~\onlinecite{shishkin2007self}) & Exp.\tabularnewline \hline
\hline Si & 0.446 & 0.56 & 1.29 & (0.04) & 1.35 & (0.04) & 1.20{*} &
1.28{*} & 1.3\footnote{Ref.~\onlinecite{tiago2004effect}}
(1.17{*})$^{a}$\tabularnewline \hline SiC & 0.293 & 1.37 & 2.29 &
(0.04) & 2.35 & (0.04) & 2.43 & 2.64 &
2.42 \footnote{Ref.~\onlinecite{bimberg1981calculation}}\tabularnewline
\hline AlP & 0.368 & 1.46 & 2.41 & (0.03) & 2.50 & (0.03) & 2.59 &
2.77 & 2.52 $^{c}$\tabularnewline \hline C & 0.336 & 4.16 & 5.40 &
(0.06) & 5.47 & (0.06) & 5.50 & 5.99 & 5.48 $^{d}$\tabularnewline
\hline BN & 0.380 & 4.48 & 6.21 & (0.06) & 6.41 & (0.07) & 6.10 & 6.73
& 6.1 - 6.4 $^{e}$\tabularnewline \hline
\end{tabular}

\caption{Fundamental band gaps ($E_{g}$) for a sample of solids. For
  each system a $4\times4\times4$ conventional super-cell (with 2048
  valence electrons) was used. Values labeled by {*} are for the
  minimum fundamental gap, which is not accessed in our calculations
  (see text). }

\label{tab_solids}
\end{table*}

\section{\label{sec:Summary-and-Conclusions}Summary and Conclusions}

In this paper we derived a general form of Green's function
self-consistency in the time domain and introduced its simplified
form, which we label $\bar{\Delta}GW_{0}$. The underlying assumption
of our method is that the differences between Kohn-Sham eigenvalues
and quasiparticle energies are approximately just two constants, one
for occupied and one for unoccupied states. We approximate this
scissors-like correction by the corrections to the HOMO and LUMO
energies.

Our approach is merely an \emph{a-posteriori} treatment of the
time-dependent self-energy matrix. Hence, $\bar{\Delta}GW_{0}$ has
essentially no additional computational cost beyond that of a one-shot
$G_{0}W_{0}$ calculation. In conjunction with the nearly linear
scaling stochastic $G_{0}W_{0}$, it is easily applicable to extremely
large systems with thousands of electrons. The combined method is best
labeled as stochastic $\bar{\Delta}GW_{0}$ or just abbreviated as
stochastic $GW_{0}$.

We tested stochastic $\bar{\Delta}GW_{0}$ on molecules and on periodic
semiconductors and insulators with large periodic supercells
containing 2048 electrons. The predicted ionization potentials and
fundamental band gaps are overall much better than one-shot
$G_{0}W_{0}$ values when compared to high-level methods and/or
experiments. Our simplified self-consistency treatment is especially
appropriate for large molecules and periodic systems, and for the
latter it yields a mean absolute error of only 0.03~eV.

The stochastic partially self-consistent $\bar{\Delta}GW_{0}$ approach
presented here is both accurate and efficient, opening the door to
many future applications in chemistry, physics nano- and material
sciences.
\begin{acknowledgments}
V.V., E.R. and D.N. were supported by the Center for Computational
Study of Excited State Phenomena in Energy Materials (C2SEPEM) at the
Lawrence Berkeley National Laboratory, which is funded by the
U.S. Department of Energy, Office of Science, Basic Energy Sciences,
Materials Sciences and Engineering Division under contract
No. DEAC02-05CH11231 as part of the Computational Materials Sciences
Program. R.B. is grateful for support by the Binational Science
Foundation, Grant 2015687 and for support from the Israel Science
Foundation \textendash{} FIRST Program, Grant No. 1700/14. The
calculations were performed as part of the XSEDE computational Project
No. TG-CHE170058 \cite{towns2014xsede}.
\end{acknowledgments}

\bibliographystyle{apsrev4-1}

\begin{thebibliography}{65}%
\makeatletter
\providecommand \@ifxundefined [1]{%
 \@ifx{#1\undefined}
}%
\providecommand \@ifnum [1]{%
 \ifnum #1\expandafter \@firstoftwo
 \else \expandafter \@secondoftwo
 \fi
}%
\providecommand \@ifx [1]{%
 \ifx #1\expandafter \@firstoftwo
 \else \expandafter \@secondoftwo
 \fi
}%
\providecommand \natexlab [1]{#1}%
\providecommand \enquote  [1]{``#1''}%
\providecommand \bibnamefont  [1]{#1}%
\providecommand \bibfnamefont [1]{#1}%
\providecommand \citenamefont [1]{#1}%
\providecommand \href@noop [0]{\@secondoftwo}%
\providecommand \href [0]{\begingroup \@sanitize@url \@href}%
\providecommand \@href[1]{\@@startlink{#1}\@@href}%
\providecommand \@@href[1]{\endgroup#1\@@endlink}%
\providecommand \@sanitize@url [0]{\catcode `\\12\catcode `\$12\catcode
  `\&12\catcode `\#12\catcode `\^12\catcode `\_12\catcode `\%12\relax}%
\providecommand \@@startlink[1]{}%
\providecommand \@@endlink[0]{}%
\providecommand \url  [0]{\begingroup\@sanitize@url \@url }%
\providecommand \@url [1]{\endgroup\@href {#1}{\urlprefix }}%
\providecommand \urlprefix  [0]{URL }%
\providecommand \Eprint [0]{\href }%
\providecommand \doibase [0]{http://dx.doi.org/}%
\providecommand \selectlanguage [0]{\@gobble}%
\providecommand \bibinfo  [0]{\@secondoftwo}%
\providecommand \bibfield  [0]{\@secondoftwo}%
\providecommand \translation [1]{[#1]}%
\providecommand \BibitemOpen [0]{}%
\providecommand \bibitemStop [0]{}%
\providecommand \bibitemNoStop [0]{.\EOS\space}%
\providecommand \EOS [0]{\spacefactor3000\relax}%
\providecommand \BibitemShut  [1]{\csname bibitem#1\endcsname}%
\let\auto@bib@innerbib\@empty
\bibitem [{\citenamefont {Hedin}(1965)}]{Hedin1965}%
  \BibitemOpen
  \bibfield  {author} {\bibinfo {author} {\bibfnamefont {L.}~\bibnamefont
  {Hedin}},\ }\href@noop {} {\bibfield  {journal} {\bibinfo  {journal} {Phys.
  Rev.}\ }\textbf {\bibinfo {volume} {139}},\ \bibinfo {pages} {A796} (\bibinfo
  {year} {1965})}\BibitemShut {NoStop}%
\bibitem [{\citenamefont {Aryasetiawan}\ and\ \citenamefont
  {Gunnarsson}(1998)}]{Aryasetiawan1998}%
  \BibitemOpen
  \bibfield  {author} {\bibinfo {author} {\bibfnamefont {F.}~\bibnamefont
  {Aryasetiawan}}\ and\ \bibinfo {author} {\bibfnamefont {O.}~\bibnamefont
  {Gunnarsson}},\ }\href@noop {} {\bibfield  {journal} {\bibinfo  {journal}
  {Rep. Prog. Phys.}\ }\textbf {\bibinfo {volume} {61}},\ \bibinfo {pages}
  {237} (\bibinfo {year} {1998})}\BibitemShut {NoStop}%
\bibitem [{\citenamefont {Rieger}\ \emph {et~al.}(1999)\citenamefont {Rieger},
  \citenamefont {Steinbeck}, \citenamefont {White}, \citenamefont {Rojas},\
  and\ \citenamefont {Godby}}]{Rieger1999}%
  \BibitemOpen
  \bibfield  {author} {\bibinfo {author} {\bibfnamefont {M.~M.}\ \bibnamefont
  {Rieger}}, \bibinfo {author} {\bibfnamefont {L.}~\bibnamefont {Steinbeck}},
  \bibinfo {author} {\bibfnamefont {I.}~\bibnamefont {White}}, \bibinfo
  {author} {\bibfnamefont {H.}~\bibnamefont {Rojas}}, \ and\ \bibinfo {author}
  {\bibfnamefont {R.}~\bibnamefont {Godby}},\ }\href
  {http://www.sciencedirect.com/science/article/pii/S001046559800174X}
  {\bibfield  {journal} {\bibinfo  {journal} {Comput. Phys. Commun.}\ }\textbf
  {\bibinfo {volume} {117}},\ \bibinfo {pages} {211} (\bibinfo {year}
  {1999})}\BibitemShut {NoStop}%
\bibitem [{\citenamefont {Steinbeck}\ \emph {et~al.}(1999)\citenamefont
  {Steinbeck}, \citenamefont {Rubio}, \citenamefont {Reining}, \citenamefont
  {Torrent}, \citenamefont {White},\ and\ \citenamefont
  {Godby}}]{Steinbeck1999}%
  \BibitemOpen
  \bibfield  {author} {\bibinfo {author} {\bibfnamefont {L.}~\bibnamefont
  {Steinbeck}}, \bibinfo {author} {\bibfnamefont {A.}~\bibnamefont {Rubio}},
  \bibinfo {author} {\bibfnamefont {L.}~\bibnamefont {Reining}}, \bibinfo
  {author} {\bibfnamefont {M.}~\bibnamefont {Torrent}}, \bibinfo {author}
  {\bibfnamefont {I.}~\bibnamefont {White}}, \ and\ \bibinfo {author}
  {\bibfnamefont {R.}~\bibnamefont {Godby}},\ }\href
  {http://arxiv.org/abs/cond-mat/9908372} {\bibfield  {journal} {\bibinfo
  {journal} {Comput. Phys. Commun.}\ }\textbf {\bibinfo {volume} {125}},\
  \bibinfo {pages} {05} (\bibinfo {year} {1999})}\BibitemShut {NoStop}%
\bibitem [{\citenamefont {Onida}\ \emph {et~al.}(2002)\citenamefont {Onida},
  \citenamefont {Reining},\ and\ \citenamefont {Rubio}}]{Onida2002}%
  \BibitemOpen
  \bibfield  {author} {\bibinfo {author} {\bibfnamefont {G.}~\bibnamefont
  {Onida}}, \bibinfo {author} {\bibfnamefont {L.}~\bibnamefont {Reining}}, \
  and\ \bibinfo {author} {\bibfnamefont {A.}~\bibnamefont {Rubio}},\
  }\href@noop {} {\bibfield  {journal} {\bibinfo  {journal} {Rev. Mod. Phys.}\
  }\textbf {\bibinfo {volume} {74}},\ \bibinfo {pages} {601} (\bibinfo {year}
  {2002})}\BibitemShut {NoStop}%
\bibitem [{\citenamefont {Rubio}\ and\ \citenamefont
  {Louie}(2005)}]{Rubio2005}%
  \BibitemOpen
  \bibfield  {author} {\bibinfo {author} {\bibfnamefont {A.}~\bibnamefont
  {Rubio}}\ and\ \bibinfo {author} {\bibfnamefont {S.~G.}\ \bibnamefont
  {Louie}},\ }\enquote {\bibinfo {title} {Quasiparticle {AND} optical
  properties of solids {AND} nanostructures: The gw-bse approach},}\ in\
  \href@noop {} {\emph {\bibinfo {booktitle} {Handbook of materials
  modeling}}},\ \bibinfo {editor} {edited by\ \bibinfo {editor} {\bibfnamefont
  {S.}~\bibnamefont {Yip}}}\ (\bibinfo  {publisher} {Springer},\ \bibinfo
  {address} {Dordrecht ; New York},\ \bibinfo {year} {2005})\ p.\ \bibinfo
  {pages} {215}\BibitemShut {NoStop}%
\bibitem [{\citenamefont {Friedrich}\ and\ \citenamefont
  {Schindlmayr}(2006)}]{Friedrich2006}%
  \BibitemOpen
  \bibfield  {author} {\bibinfo {author} {\bibfnamefont {C.}~\bibnamefont
  {Friedrich}}\ and\ \bibinfo {author} {\bibfnamefont {A.}~\bibnamefont
  {Schindlmayr}},\ }\href@noop {} {\bibfield  {journal} {\bibinfo  {journal}
  {NIC Series}\ }\textbf {\bibinfo {volume} {31}},\ \bibinfo {pages} {335}
  (\bibinfo {year} {2006})}\BibitemShut {NoStop}%
\bibitem [{\citenamefont {Shishkin}\ and\ \citenamefont
  {Kresse}(2007{\natexlab{a}})}]{Shishkin2007}%
  \BibitemOpen
  \bibfield  {author} {\bibinfo {author} {\bibfnamefont {M.}~\bibnamefont
  {Shishkin}}\ and\ \bibinfo {author} {\bibfnamefont {G.}~\bibnamefont
  {Kresse}},\ }\href {http://prb.aps.org/abstract/PRB/v75/i23/e235102}
  {\bibfield  {journal} {\bibinfo  {journal} {Phys. Rev. B}\ }\textbf {\bibinfo
  {volume} {75}},\ \bibinfo {pages} {235102} (\bibinfo {year}
  {2007}{\natexlab{a}})}\BibitemShut {NoStop}%
\bibitem [{\citenamefont {Trevisanutto}\ \emph {et~al.}(2008)\citenamefont
  {Trevisanutto}, \citenamefont {Giorgetti}, \citenamefont {Reining},
  \citenamefont {Ladisa},\ and\ \citenamefont {Olevano}}]{Trevisanutto2008}%
  \BibitemOpen
  \bibfield  {author} {\bibinfo {author} {\bibfnamefont {P.~E.}\ \bibnamefont
  {Trevisanutto}}, \bibinfo {author} {\bibfnamefont {C.}~\bibnamefont
  {Giorgetti}}, \bibinfo {author} {\bibfnamefont {L.}~\bibnamefont {Reining}},
  \bibinfo {author} {\bibfnamefont {M.}~\bibnamefont {Ladisa}}, \ and\ \bibinfo
  {author} {\bibfnamefont {V.}~\bibnamefont {Olevano}},\ }\href
  {http://prl.aps.org/abstract/PRL/v101/i22/e226405} {\bibfield  {journal}
  {\bibinfo  {journal} {Phys. Rev. Lett.}\ }\textbf {\bibinfo {volume} {101}},\
  \bibinfo {pages} {226405} (\bibinfo {year} {2008})}\BibitemShut {NoStop}%
\bibitem [{\citenamefont {Rostgaard}\ \emph
  {et~al.}(2010{\natexlab{a}})\citenamefont {Rostgaard}, \citenamefont
  {Jacobsen},\ and\ \citenamefont {Thygesen}}]{Rostgaard2010}%
  \BibitemOpen
  \bibfield  {author} {\bibinfo {author} {\bibfnamefont {C.}~\bibnamefont
  {Rostgaard}}, \bibinfo {author} {\bibfnamefont {K.~W.}\ \bibnamefont
  {Jacobsen}}, \ and\ \bibinfo {author} {\bibfnamefont {K.~S.}\ \bibnamefont
  {Thygesen}},\ }\href {http://prb.aps.org/abstract/PRB/v81/i8/e085103}
  {\bibfield  {journal} {\bibinfo  {journal} {Phys. Rev. B}\ }\textbf {\bibinfo
  {volume} {81}},\ \bibinfo {pages} {085103} (\bibinfo {year}
  {2010}{\natexlab{a}})}\BibitemShut {NoStop}%
\bibitem [{\citenamefont {Tamblyn}\ \emph {et~al.}(2011)\citenamefont
  {Tamblyn}, \citenamefont {Darancet}, \citenamefont {Quek}, \citenamefont
  {Bonev},\ and\ \citenamefont {Neaton}}]{Tamblyn2011}%
  \BibitemOpen
  \bibfield  {author} {\bibinfo {author} {\bibfnamefont {I.}~\bibnamefont
  {Tamblyn}}, \bibinfo {author} {\bibfnamefont {P.}~\bibnamefont {Darancet}},
  \bibinfo {author} {\bibfnamefont {S.~Y.}\ \bibnamefont {Quek}}, \bibinfo
  {author} {\bibfnamefont {S.~A.}\ \bibnamefont {Bonev}}, \ and\ \bibinfo
  {author} {\bibfnamefont {J.~B.}\ \bibnamefont {Neaton}},\ }\href
  {http://prb.aps.org/abstract/PRB/v84/i20/e201402} {\bibfield  {journal}
  {\bibinfo  {journal} {Phys. Rev. B}\ }\textbf {\bibinfo {volume} {84}},\
  \bibinfo {pages} {201402} (\bibinfo {year} {2011})}\BibitemShut {NoStop}%
\bibitem [{\citenamefont {van Setten}\ \emph {et~al.}(2012)\citenamefont {van
  Setten}, \citenamefont {Weigend},\ and\ \citenamefont {Evers}}]{Setten2012}%
  \BibitemOpen
  \bibfield  {author} {\bibinfo {author} {\bibfnamefont {M.}~\bibnamefont {van
  Setten}}, \bibinfo {author} {\bibfnamefont {F.}~\bibnamefont {Weigend}}, \
  and\ \bibinfo {author} {\bibfnamefont {F.}~\bibnamefont {Evers}},\ }\href
  {http://pubs.acs.org/doi/abs/10.1021/ct300648t} {\bibfield  {journal}
  {\bibinfo  {journal} {J. Chem. Theory Comput.}\ }\textbf {\bibinfo {volume}
  {9}},\ \bibinfo {pages} {232} (\bibinfo {year} {2012})}\BibitemShut {NoStop}%
\bibitem [{\citenamefont {Stefanucci}\ and\ \citenamefont {van
  Leeuwen}(2013)}]{Stefanucci2013}%
  \BibitemOpen
  \bibfield  {author} {\bibinfo {author} {\bibfnamefont {G.}~\bibnamefont
  {Stefanucci}}\ and\ \bibinfo {author} {\bibfnamefont {R.}~\bibnamefont {van
  Leeuwen}},\ }\href@noop {} {\emph {\bibinfo {title} {Nonequilibrium Many-Body
  Theory of Quantum Systems: A Modern Introduction}}}\ (\bibinfo  {publisher}
  {Cambridge University Press},\ \bibinfo {year} {2013})\BibitemShut {NoStop}%
\bibitem [{\citenamefont {Govoni}\ and\ \citenamefont
  {Galli}(2015)}]{Govoni2015}%
  \BibitemOpen
  \bibfield  {author} {\bibinfo {author} {\bibfnamefont {M.}~\bibnamefont
  {Govoni}}\ and\ \bibinfo {author} {\bibfnamefont {G.}~\bibnamefont {Galli}},\
  }\href@noop {} {\bibfield  {journal} {\bibinfo  {journal} {J. Chem. Theory Comput.}\ }\textbf {\bibinfo {volume} {11}},\
  \bibinfo {pages} {2680} (\bibinfo {year} {2015})}\BibitemShut {NoStop}%
\bibitem [{\citenamefont {Kaplan}\ \emph
  {et~al.}(2016{\natexlab{a}})\citenamefont {Kaplan}, \citenamefont {Harding},
  \citenamefont {Seiler}, \citenamefont {Weigend}, \citenamefont {Evers},\ and\
  \citenamefont {van Setten}}]{Kaplan2016}%
  \BibitemOpen
  \bibfield  {author} {\bibinfo {author} {\bibfnamefont {F.}~\bibnamefont
  {Kaplan}}, \bibinfo {author} {\bibfnamefont {M.~E.}\ \bibnamefont {Harding}},
  \bibinfo {author} {\bibfnamefont {C.}~\bibnamefont {Seiler}}, \bibinfo
  {author} {\bibfnamefont {F.}~\bibnamefont {Weigend}}, \bibinfo {author}
  {\bibfnamefont {F.}~\bibnamefont {Evers}}, \ and\ \bibinfo {author}
  {\bibfnamefont {M.~J.}\ \bibnamefont {van Setten}},\ }\href@noop {}
  {\bibfield  {journal} {\bibinfo  {journal} {J. Chem. Theory Comput.}\ }
  (\bibinfo {year} {2016}{\natexlab{a}})}\BibitemShut {NoStop}%
\bibitem [{\citenamefont {Hybertsen}\ and\ \citenamefont
  {Louie}(1985)}]{Hybertsen1985}%
  \BibitemOpen
  \bibfield  {author} {\bibinfo {author} {\bibfnamefont {M.~S.}\ \bibnamefont
  {Hybertsen}}\ and\ \bibinfo {author} {\bibfnamefont {S.~G.}\ \bibnamefont
  {Louie}},\ }\href@noop {} {\bibfield  {journal} {\bibinfo  {journal} {Phys.
  Rev. Lett.}\ }\textbf {\bibinfo {volume} {55}},\ \bibinfo {pages} {1418}
  (\bibinfo {year} {1985})}\BibitemShut {NoStop}%
\bibitem [{\citenamefont {Hybertsen}\ and\ \citenamefont
  {Louie}(1986{\natexlab{a}})}]{Hybertsen1986}%
  \BibitemOpen
  \bibfield  {author} {\bibinfo {author} {\bibfnamefont {M.~S.}\ \bibnamefont
  {Hybertsen}}\ and\ \bibinfo {author} {\bibfnamefont {S.~G.}\ \bibnamefont
  {Louie}},\ }\href@noop {} {\bibfield  {journal} {\bibinfo  {journal} {Phys.
  Rev. B}\ }\textbf {\bibinfo {volume} {34}},\ \bibinfo {pages} {5390}
  (\bibinfo {year} {1986}{\natexlab{a}})}\BibitemShut {NoStop}%
\bibitem [{\citenamefont {Kohn}\ and\ \citenamefont {Sham}(1965)}]{Kohn1965}%
  \BibitemOpen
  \bibfield  {author} {\bibinfo {author} {\bibfnamefont {W.}~\bibnamefont
  {Kohn}}\ and\ \bibinfo {author} {\bibfnamefont {L.~J.}\ \bibnamefont
  {Sham}},\ }\href@noop {} {\bibfield  {journal} {\bibinfo  {journal} {Phys.
  Rev.}\ }\textbf {\bibinfo {volume} {140}},\ \bibinfo {pages} {A1133}
  (\bibinfo {year} {1965})}\BibitemShut {NoStop}%
\bibitem [{\citenamefont {Hohenberg}\ and\ \citenamefont
  {Kohn}(1964)}]{Hohenberg1964}%
  \BibitemOpen
  \bibfield  {author} {\bibinfo {author} {\bibfnamefont {P.}~\bibnamefont
  {Hohenberg}}\ and\ \bibinfo {author} {\bibfnamefont {W.}~\bibnamefont
  {Kohn}},\ }\href@noop {} {\bibfield  {journal} {\bibinfo  {journal} {Phys.
  Rev.}\ }\textbf {\bibinfo {volume} {136}},\ \bibinfo {pages} {B864} (\bibinfo
  {year} {1964})}\BibitemShut {NoStop}%
\bibitem [{\citenamefont {Faleev}\ \emph {et~al.}(2004)\citenamefont {Faleev},
  \citenamefont {van~Schilfgaarde},\ and\ \citenamefont {Kotani}}]{Faleev2004}%
  \BibitemOpen
  \bibfield  {author} {\bibinfo {author} {\bibfnamefont {S.~V.}\ \bibnamefont
  {Faleev}}, \bibinfo {author} {\bibfnamefont {M.}~\bibnamefont
  {van~Schilfgaarde}}, \ and\ \bibinfo {author} {\bibfnamefont
  {T.}~\bibnamefont {Kotani}},\ }\href@noop {} {\bibfield  {journal} {\bibinfo
  {journal} {Phys. Rev. Lett.}\ }\textbf {\bibinfo {volume} {93}},\ \bibinfo
  {pages} {126406} (\bibinfo {year} {2004})}\BibitemShut {NoStop}%
\bibitem [{\citenamefont {Caruso}\ \emph
  {et~al.}(2012{\natexlab{a}})\citenamefont {Caruso}, \citenamefont {Rinke},
  \citenamefont {Ren}, \citenamefont {Scheffler},\ and\ \citenamefont
  {Rubio}}]{Caruso2012}%
  \BibitemOpen
  \bibfield  {author} {\bibinfo {author} {\bibfnamefont {F.}~\bibnamefont
  {Caruso}}, \bibinfo {author} {\bibfnamefont {P.}~\bibnamefont {Rinke}},
  \bibinfo {author} {\bibfnamefont {X.}~\bibnamefont {Ren}}, \bibinfo {author}
  {\bibfnamefont {M.}~\bibnamefont {Scheffler}}, \ and\ \bibinfo {author}
  {\bibfnamefont {A.}~\bibnamefont {Rubio}},\ }\href
  {http://prb.aps.org/abstract/PRB/v86/i8/e081102} {\bibfield  {journal}
  {\bibinfo  {journal} {Phys. Rev. B}\ }\textbf {\bibinfo {volume} {86}},\
  \bibinfo {pages} {081102} (\bibinfo {year} {2012}{\natexlab{a}})}\BibitemShut
  {NoStop}%
\bibitem [{\citenamefont {Bruneval}\ and\ \citenamefont
  {Marques}(2012)}]{BrunevalMarques2012}%
  \BibitemOpen
  \bibfield  {author} {\bibinfo {author} {\bibfnamefont {F.}~\bibnamefont
  {Bruneval}}\ and\ \bibinfo {author} {\bibfnamefont {M.~A.}\ \bibnamefont
  {Marques}},\ }\href@noop {} {\bibfield  {journal} {\bibinfo  {journal} {J.
  Chem. Theory Comput.}\ }\textbf {\bibinfo {volume} {9}},\ \bibinfo {pages}
  {324} (\bibinfo {year} {2012})}\BibitemShut {NoStop}%
\bibitem [{\citenamefont {Bruneval}(2012)}]{Bruneval2012}%
  \BibitemOpen
  \bibfield  {author} {\bibinfo {author} {\bibfnamefont {F.}~\bibnamefont
  {Bruneval}},\ }\href@noop {} {\bibfield  {journal} {\bibinfo  {journal} {J. Chem. Phys.}\ }\textbf {\bibinfo {volume} {136}},\ \bibinfo
  {pages} {194107} (\bibinfo {year} {2012})}\BibitemShut {NoStop}%
\bibitem [{\citenamefont {van Setten}\ \emph
  {et~al.}(2015{\natexlab{a}})\citenamefont {van Setten}, \citenamefont
  {Caruso}, \citenamefont {Sharifzadeh}, \citenamefont {Ren}, \citenamefont
  {Scheffler}, \citenamefont {Liu}, \citenamefont {Lischner}, \citenamefont
  {Lin}, \citenamefont {Deslippe}, \citenamefont {Louie}, \citenamefont {Yang},
  \citenamefont {Weigend}, \citenamefont {Neaton}, \citenamefont {Evers},\ and\
  \citenamefont {Rinke}}]{Setten2015}%
  \BibitemOpen
  \bibfield  {author} {\bibinfo {author} {\bibfnamefont {M.~J.}\ \bibnamefont
  {van Setten}}, \bibinfo {author} {\bibfnamefont {F.}~\bibnamefont {Caruso}},
  \bibinfo {author} {\bibfnamefont {S.}~\bibnamefont {Sharifzadeh}}, \bibinfo
  {author} {\bibfnamefont {X.}~\bibnamefont {Ren}}, \bibinfo {author}
  {\bibfnamefont {M.}~\bibnamefont {Scheffler}}, \bibinfo {author}
  {\bibfnamefont {F.}~\bibnamefont {Liu}}, \bibinfo {author} {\bibfnamefont
  {J.}~\bibnamefont {Lischner}}, \bibinfo {author} {\bibfnamefont
  {L.}~\bibnamefont {Lin}}, \bibinfo {author} {\bibfnamefont {J.~R.}\
  \bibnamefont {Deslippe}}, \bibinfo {author} {\bibfnamefont {S.~G.}\
  \bibnamefont {Louie}}, \bibinfo {author} {\bibfnamefont {C.}~\bibnamefont
  {Yang}}, \bibinfo {author} {\bibfnamefont {F.}~\bibnamefont {Weigend}},
  \bibinfo {author} {\bibfnamefont {J.~B.}\ \bibnamefont {Neaton}}, \bibinfo
  {author} {\bibfnamefont {F.}~\bibnamefont {Evers}}, \ and\ \bibinfo {author}
  {\bibfnamefont {P.}~\bibnamefont {Rinke}},\ }\href@noop {} {\bibfield
  {journal} {\bibinfo  {journal} {J. Chem. Theory Comput.}\ }\textbf {\bibinfo
  {volume} {11}},\ \bibinfo {pages} {5665} (\bibinfo {year}
  {2015}{\natexlab{a}})}\BibitemShut {NoStop}%
\bibitem [{\citenamefont {Knight}\ \emph {et~al.}(2016)\citenamefont {Knight},
  \citenamefont {Wang}, \citenamefont {Gallandi}, \citenamefont
  {Dolgounitcheva}, \citenamefont {Ren}, \citenamefont {Ortiz}, \citenamefont
  {Rinke}, \citenamefont {K{\"o}rzd{\"o}rfer},\ and\ \citenamefont
  {Marom}}]{Knight2016}%
  \BibitemOpen
  \bibfield  {author} {\bibinfo {author} {\bibfnamefont {J.~W.}\ \bibnamefont
  {Knight}}, \bibinfo {author} {\bibfnamefont {X.}~\bibnamefont {Wang}},
  \bibinfo {author} {\bibfnamefont {L.}~\bibnamefont {Gallandi}}, \bibinfo
  {author} {\bibfnamefont {O.}~\bibnamefont {Dolgounitcheva}}, \bibinfo
  {author} {\bibfnamefont {X.}~\bibnamefont {Ren}}, \bibinfo {author}
  {\bibfnamefont {J.~V.}\ \bibnamefont {Ortiz}}, \bibinfo {author}
  {\bibfnamefont {P.}~\bibnamefont {Rinke}}, \bibinfo {author} {\bibfnamefont
  {T.}~\bibnamefont {K{\"o}rzd{\"o}rfer}}, \ and\ \bibinfo {author}
  {\bibfnamefont {N.}~\bibnamefont {Marom}},\ }\href@noop {} {\bibfield
  {journal} {\bibinfo  {journal} {J. Chem. Theory Comput.}\ } (\bibinfo {year}
  {2016})}\BibitemShut {NoStop}%
\bibitem [{\citenamefont {Stan}\ \emph {et~al.}(2006)\citenamefont {Stan},
  \citenamefont {Dahlen},\ and\ \citenamefont {van~Leeuwen}}]{stan2006fully}%
  \BibitemOpen
  \bibfield  {author} {\bibinfo {author} {\bibfnamefont {A.}~\bibnamefont
  {Stan}}, \bibinfo {author} {\bibfnamefont {N.~E.}\ \bibnamefont {Dahlen}}, \
  and\ \bibinfo {author} {\bibfnamefont {R.}~\bibnamefont {van~Leeuwen}},\
  }\href@noop {} {\bibfield  {journal} {\bibinfo  {journal} {Europhysics
  Lett.)}\ }\textbf {\bibinfo {volume} {76}},\ \bibinfo {pages} {298}
  (\bibinfo {year} {2006})}\BibitemShut {NoStop}%
\bibitem [{\citenamefont {Stan}\ \emph {et~al.}(2009)\citenamefont {Stan},
  \citenamefont {Dahlen},\ and\ \citenamefont
  {van~Leeuwen}}]{StanVanLeeuwen2009}%
  \BibitemOpen
  \bibfield  {author} {\bibinfo {author} {\bibfnamefont {A.}~\bibnamefont
  {Stan}}, \bibinfo {author} {\bibfnamefont {N.~E.}\ \bibnamefont {Dahlen}}, \
  and\ \bibinfo {author} {\bibfnamefont {R.}~\bibnamefont {van~Leeuwen}},\
  }\href@noop {} {\bibfield  {journal} {\bibinfo  {journal} {J. Chem. Phys.}\
  }\textbf {\bibinfo {volume} {130}},\ \bibinfo {pages} {114105} (\bibinfo
  {year} {2009})}\BibitemShut {NoStop}%
\bibitem [{\citenamefont {Rostgaard}\ \emph
  {et~al.}(2010{\natexlab{b}})\citenamefont {Rostgaard}, \citenamefont
  {Jacobsen},\ and\ \citenamefont {Thygesen}}]{RostgaardJacobsenThygesen2010}%
  \BibitemOpen
  \bibfield  {author} {\bibinfo {author} {\bibfnamefont {C.}~\bibnamefont
  {Rostgaard}}, \bibinfo {author} {\bibfnamefont {K.~W.}\ \bibnamefont
  {Jacobsen}}, \ and\ \bibinfo {author} {\bibfnamefont {K.~S.}\ \bibnamefont
  {Thygesen}},\ }\href@noop {} {\bibfield  {journal} {\bibinfo  {journal}
  {Phys. Rev. B}\ }\textbf {\bibinfo {volume} {81}},\ \bibinfo {pages} {085103}
  (\bibinfo {year} {2010}{\natexlab{b}})}\BibitemShut {NoStop}%
\bibitem [{\citenamefont {Blase}\ and\ \citenamefont
  {Attaccalite}(2011)}]{Blase2011}%
  \BibitemOpen
  \bibfield  {author} {\bibinfo {author} {\bibfnamefont {X.}~\bibnamefont
  {Blase}}\ and\ \bibinfo {author} {\bibfnamefont {C.}~\bibnamefont
  {Attaccalite}},\ }\href@noop {} {\bibfield  {journal} {\bibinfo  {journal}
  {Appl. Phys. Lett.}\ }\textbf {\bibinfo {volume} {99}},\ \bibinfo {pages}
  {171909} (\bibinfo {year} {2011})}\BibitemShut {NoStop}%
\bibitem [{\citenamefont {Deslippe}\ \emph {et~al.}(2012)\citenamefont
  {Deslippe}, \citenamefont {Samsonidze}, \citenamefont {Strubbe},
  \citenamefont {Jain}, \citenamefont {Cohen},\ and\ \citenamefont
  {Louie}}]{Deslippe2012}%
  \BibitemOpen
  \bibfield  {author} {\bibinfo {author} {\bibfnamefont {J.}~\bibnamefont
  {Deslippe}}, \bibinfo {author} {\bibfnamefont {G.}~\bibnamefont
  {Samsonidze}}, \bibinfo {author} {\bibfnamefont {D.~A.}\ \bibnamefont
  {Strubbe}}, \bibinfo {author} {\bibfnamefont {M.}~\bibnamefont {Jain}},
  \bibinfo {author} {\bibfnamefont {M.~L.}\ \bibnamefont {Cohen}}, \ and\
  \bibinfo {author} {\bibfnamefont {S.~G.}\ \bibnamefont {Louie}},\ }\href
  {\doibase http://dx.doi.org/10.1016/j.cpc.2011.12.006} {\bibfield  {journal}
  {\bibinfo  {journal} {Comput. Phys. Commun.}\ }\textbf {\bibinfo {volume}
  {183}},\ \bibinfo {pages} {1269 } (\bibinfo {year} {2012})}\BibitemShut
  {NoStop}%
\bibitem [{\citenamefont {Nguyen}\ \emph {et~al.}(2012)\citenamefont {Nguyen},
  \citenamefont {Pham}, \citenamefont {Rocca},\ and\ \citenamefont
  {Galli}}]{Nguyen2012}%
  \BibitemOpen
  \bibfield  {author} {\bibinfo {author} {\bibfnamefont {H.-V.}\ \bibnamefont
  {Nguyen}}, \bibinfo {author} {\bibfnamefont {T.~A.}\ \bibnamefont {Pham}},
  \bibinfo {author} {\bibfnamefont {D.}~\bibnamefont {Rocca}}, \ and\ \bibinfo
  {author} {\bibfnamefont {G.}~\bibnamefont {Galli}},\ }\href
  {http://prb.aps.org/abstract/PRB/v85/i8/e081101} {\bibfield  {journal}
  {\bibinfo  {journal} {Phys. Rev. B}\ }\textbf {\bibinfo {volume} {85}},\
  \bibinfo {pages} {081101} (\bibinfo {year} {2012})}\BibitemShut {NoStop}%
\bibitem [{\citenamefont {Caruso}\ \emph
  {et~al.}(2012{\natexlab{b}})\citenamefont {Caruso}, \citenamefont {Rinke},
  \citenamefont {Ren}, \citenamefont {Scheffler},\ and\ \citenamefont
  {Rubio}}]{caruso2012unified}%
  \BibitemOpen
  \bibfield  {author} {\bibinfo {author} {\bibfnamefont {F.}~\bibnamefont
  {Caruso}}, \bibinfo {author} {\bibfnamefont {P.}~\bibnamefont {Rinke}},
  \bibinfo {author} {\bibfnamefont {X.}~\bibnamefont {Ren}}, \bibinfo {author}
  {\bibfnamefont {M.}~\bibnamefont {Scheffler}}, \ and\ \bibinfo {author}
  {\bibfnamefont {A.}~\bibnamefont {Rubio}},\ }\href@noop {} {\bibfield
  {journal} {\bibinfo  {journal} {Phys. Rev. B}\ }\textbf {\bibinfo {volume}
  {86}},\ \bibinfo {pages} {081102} (\bibinfo {year}
  {2012}{\natexlab{b}})}\BibitemShut {NoStop}%
\bibitem [{\citenamefont {Caruso}\ \emph {et~al.}(2013)\citenamefont {Caruso},
  \citenamefont {Rinke}, \citenamefont {Ren}, \citenamefont {Rubio},\ and\
  \citenamefont {Scheffler}}]{caruso2013self}%
  \BibitemOpen
  \bibfield  {author} {\bibinfo {author} {\bibfnamefont {F.}~\bibnamefont
  {Caruso}}, \bibinfo {author} {\bibfnamefont {P.}~\bibnamefont {Rinke}},
  \bibinfo {author} {\bibfnamefont {X.}~\bibnamefont {Ren}}, \bibinfo {author}
  {\bibfnamefont {A.}~\bibnamefont {Rubio}}, \ and\ \bibinfo {author}
  {\bibfnamefont {M.}~\bibnamefont {Scheffler}},\ }\href@noop {} {\bibfield
  {journal} {\bibinfo  {journal} {Phys. Rev. B}\ }\textbf {\bibinfo {volume}
  {88}},\ \bibinfo {pages} {075105} (\bibinfo {year} {2013})}\BibitemShut
  {NoStop}%
\bibitem [{\citenamefont {Koval}\ \emph {et~al.}(2014)\citenamefont {Koval},
  \citenamefont {Foerster},\ and\ \citenamefont
  {S{\'a}nchez-Portal}}]{koval2014fully}%
  \BibitemOpen
  \bibfield  {author} {\bibinfo {author} {\bibfnamefont {P.}~\bibnamefont
  {Koval}}, \bibinfo {author} {\bibfnamefont {D.}~\bibnamefont {Foerster}}, \
  and\ \bibinfo {author} {\bibfnamefont {D.}~\bibnamefont
  {S{\'a}nchez-Portal}},\ }\href@noop {} {\bibfield  {journal} {\bibinfo
  {journal} {Phys. Rev. B}\ }\textbf {\bibinfo {volume} {89}},\ \bibinfo
  {pages} {155417} (\bibinfo {year} {2014})}\BibitemShut {NoStop}%
\bibitem [{\citenamefont {Wang}(2015)}]{wang2015fully}%
  \BibitemOpen
  \bibfield  {author} {\bibinfo {author} {\bibfnamefont {L.-W.}\ \bibnamefont
  {Wang}},\ }\href@noop {} {\bibfield  {journal} {\bibinfo  {journal} {Phys.
  Rev. B}\ }\textbf {\bibinfo {volume} {91}},\ \bibinfo {pages} {125135}
  (\bibinfo {year} {2015})}\BibitemShut {NoStop}%
\bibitem [{\citenamefont {Holm}\ and\ \citenamefont {von
  Barth}(1998)}]{holm1998fully}%
  \BibitemOpen
  \bibfield  {author} {\bibinfo {author} {\bibfnamefont {B.}~\bibnamefont
  {Holm}}\ and\ \bibinfo {author} {\bibfnamefont {U.}~\bibnamefont {von
  Barth}},\ }\href@noop {} {\bibfield  {journal} {\bibinfo  {journal} {Phys.
  Rev. B}\ }\textbf {\bibinfo {volume} {57}},\ \bibinfo {pages} {2108}
  (\bibinfo {year} {1998})}\BibitemShut {NoStop}%
\bibitem [{\citenamefont {Bruneval}\ and\ \citenamefont
  {Gatti}(2014)}]{BrunevalGatti2014}%
  \BibitemOpen
  \bibfield  {author} {\bibinfo {author} {\bibfnamefont {F.}~\bibnamefont
  {Bruneval}}\ and\ \bibinfo {author} {\bibfnamefont {M.}~\bibnamefont
  {Gatti}},\ }in\ \href@noop {} {\emph {\bibinfo {booktitle} {First Principles
  Approaches to Spectroscopic Properties of Complex Materials}}}\ (\bibinfo
  {publisher} {Springer},\ \bibinfo {year} {2014})\ pp.\ \bibinfo {pages}
  {99--135}\BibitemShut {NoStop}%
\bibitem [{\citenamefont {van Schilfgaarde}\ \emph {et~al.}(2006)\citenamefont
  {van Schilfgaarde}, \citenamefont {Kotani},\ and\ \citenamefont
  {Faleev}}]{van2006quasiparticle}%
  \BibitemOpen
  \bibfield  {author} {\bibinfo {author} {\bibfnamefont {M.}~\bibnamefont {van
  Schilfgaarde}}, \bibinfo {author} {\bibfnamefont {T.}~\bibnamefont {Kotani}},
  \ and\ \bibinfo {author} {\bibfnamefont {S.}~\bibnamefont {Faleev}},\
  }\href@noop {} {\bibfield  {journal} {\bibinfo  {journal} {Phys. Rev. Lett.}\
  }\textbf {\bibinfo {volume} {96}},\ \bibinfo {pages} {226402} (\bibinfo
  {year} {2006})}\BibitemShut {NoStop}%
\bibitem [{\citenamefont {Kotani}\ \emph {et~al.}(2007)\citenamefont {Kotani},
  \citenamefont {van Schilfgaarde},\ and\ \citenamefont
  {Faleev}}]{kotani2007quasiparticle}%
  \BibitemOpen
  \bibfield  {author} {\bibinfo {author} {\bibfnamefont {T.}~\bibnamefont
  {Kotani}}, \bibinfo {author} {\bibfnamefont {M.}~\bibnamefont {van
  Schilfgaarde}}, \ and\ \bibinfo {author} {\bibfnamefont {S.~V.}\ \bibnamefont
  {Faleev}},\ }\href@noop {} {\bibfield  {journal} {\bibinfo  {journal} {Phys.
  Rev. B}\ }\textbf {\bibinfo {volume} {76}},\ \bibinfo {pages} {165106}
  (\bibinfo {year} {2007})}\BibitemShut {NoStop}%
\bibitem [{\citenamefont {Kaplan}\ \emph {et~al.}(2015)\citenamefont {Kaplan},
  \citenamefont {Weigend}, \citenamefont {Evers},\ and\ \citenamefont {van
  Setten}}]{kaplan2015off}%
  \BibitemOpen
  \bibfield  {author} {\bibinfo {author} {\bibfnamefont {F.}~\bibnamefont
  {Kaplan}}, \bibinfo {author} {\bibfnamefont {F.}~\bibnamefont {Weigend}},
  \bibinfo {author} {\bibfnamefont {F.}~\bibnamefont {Evers}}, \ and\ \bibinfo
  {author} {\bibfnamefont {M.}~\bibnamefont {van Setten}},\ }\href@noop {}
  {\bibfield  {journal} {\bibinfo  {journal} {J. Chem. Theory Comput.}\ }\textbf
  {\bibinfo {volume} {11}},\ \bibinfo {pages} {5152} (\bibinfo {year}
  {2015})}\BibitemShut {NoStop}%
\bibitem [{\citenamefont {Kaplan}\ \emph
  {et~al.}(2016{\natexlab{b}})\citenamefont {Kaplan}, \citenamefont {Harding},
  \citenamefont {Seiler}, \citenamefont {Weigend}, \citenamefont {Evers},\ and\
  \citenamefont {van Setten}}]{kaplan2016quasi}%
  \BibitemOpen
  \bibfield  {author} {\bibinfo {author} {\bibfnamefont {F.}~\bibnamefont
  {Kaplan}}, \bibinfo {author} {\bibfnamefont {M.}~\bibnamefont {Harding}},
  \bibinfo {author} {\bibfnamefont {C.}~\bibnamefont {Seiler}}, \bibinfo
  {author} {\bibfnamefont {F.}~\bibnamefont {Weigend}}, \bibinfo {author}
  {\bibfnamefont {F.}~\bibnamefont {Evers}}, \ and\ \bibinfo {author}
  {\bibfnamefont {M.}~\bibnamefont {van Setten}},\ }\href@noop {} {\bibfield
  {journal} {\bibinfo  {journal} {J. Chem. Theory Comput.}\ }\textbf {\bibinfo
  {volume} {12}},\ \bibinfo {pages} {2528} (\bibinfo {year}
  {2016}{\natexlab{b}})}\BibitemShut {NoStop}%
\bibitem [{\citenamefont {Caruso}\ \emph {et~al.}(2016)\citenamefont {Caruso},
  \citenamefont {Dauth}, \citenamefont {van Setten},\ and\ \citenamefont
  {Rinke}}]{caruso2016benchmark}%
  \BibitemOpen
  \bibfield  {author} {\bibinfo {author} {\bibfnamefont {F.}~\bibnamefont
  {Caruso}}, \bibinfo {author} {\bibfnamefont {M.}~\bibnamefont {Dauth}},
  \bibinfo {author} {\bibfnamefont {M.~J.}\ \bibnamefont {van Setten}}, \ and\
  \bibinfo {author} {\bibfnamefont {P.}~\bibnamefont {Rinke}},\ }\href@noop {}
  {\bibfield  {journal} {\bibinfo  {journal} {J. Chem. Theory. Comput.}\ }\textbf
  {\bibinfo {volume} {12}},\ \bibinfo {pages} {5076} (\bibinfo {year}
  {2016})}\BibitemShut {NoStop}%
\bibitem [{\citenamefont {Shishkin}\ and\ \citenamefont
  {Kresse}(2007{\natexlab{b}})}]{shishkin2007self}%
  \BibitemOpen
  \bibfield  {author} {\bibinfo {author} {\bibfnamefont {M.}~\bibnamefont
  {Shishkin}}\ and\ \bibinfo {author} {\bibfnamefont {G.}~\bibnamefont
  {Kresse}},\ }\href@noop {} {\bibfield  {journal} {\bibinfo  {journal}
  {Phys. Rev. B}\ }\textbf {\bibinfo {volume} {75}},\ \bibinfo {pages}
  {235102} (\bibinfo {year} {2007}{\natexlab{b}})}\BibitemShut {NoStop}%
\bibitem [{\citenamefont {Northrup}\ \emph {et~al.}(1987)\citenamefont
  {Northrup}, \citenamefont {Hybertsen},\ and\ \citenamefont
  {Louie}}]{Northrup1987}%
  \BibitemOpen
  \bibfield  {author} {\bibinfo {author} {\bibfnamefont {J.~E.}\ \bibnamefont
  {Northrup}}, \bibinfo {author} {\bibfnamefont {M.~S.}\ \bibnamefont
  {Hybertsen}}, \ and\ \bibinfo {author} {\bibfnamefont {S.~G.}\ \bibnamefont
  {Louie}},\ }\href@noop {} {\bibfield  {journal} {\bibinfo  {journal}
  {Phys. Rev. Lett.}\ }\textbf {\bibinfo {volume} {59}},\ \bibinfo
  {pages} {819} (\bibinfo {year} {1987})}\BibitemShut {NoStop}%
\bibitem [{\citenamefont {Neuhauser}\ \emph {et~al.}(2014)\citenamefont
  {Neuhauser}, \citenamefont {Gao}, \citenamefont {Arntsen}, \citenamefont
  {Karshenas}, \citenamefont {Rabani},\ and\ \citenamefont
  {Baer}}]{Neuhauser2014}%
  \BibitemOpen
  \bibfield  {author} {\bibinfo {author} {\bibfnamefont {D.}~\bibnamefont
  {Neuhauser}}, \bibinfo {author} {\bibfnamefont {Y.}~\bibnamefont {Gao}},
  \bibinfo {author} {\bibfnamefont {C.}~\bibnamefont {Arntsen}}, \bibinfo
  {author} {\bibfnamefont {C.}~\bibnamefont {Karshenas}}, \bibinfo {author}
  {\bibfnamefont {E.}~\bibnamefont {Rabani}}, \ and\ \bibinfo {author}
  {\bibfnamefont {R.}~\bibnamefont {Baer}},\ }\href@noop {} {\bibfield
  {journal} {\bibinfo  {journal} {Phys. Rev. Lett.}\ }\textbf {\bibinfo
  {volume} {113}},\ \bibinfo {pages} {076402} (\bibinfo {year}
  {2014})}\BibitemShut {NoStop}%
\bibitem [{\citenamefont {Vlcek}\ \emph {et~al.}(2017)\citenamefont {Vlcek},
  \citenamefont {Rabani}, \citenamefont {Neuhauser},\ and\ \citenamefont
  {Baer}}]{vlcek2017stochastic}%
  \BibitemOpen
  \bibfield  {author} {\bibinfo {author} {\bibfnamefont {V.}~\bibnamefont
  {Vlcek}}, \bibinfo {author} {\bibfnamefont {E.}~\bibnamefont {Rabani}},
  \bibinfo {author} {\bibfnamefont {D.}~\bibnamefont {Neuhauser}}, \ and\
  \bibinfo {author} {\bibfnamefont {R.}~\bibnamefont {Baer}},\ }\href@noop {}
  {\bibfield  {journal} {\bibinfo  {journal} {J. Chem. Theory Comput.}\ }\textbf {\bibinfo {volume} {13}},\ \bibinfo {pages} {4997}
  (\bibinfo {year} {2017})}\BibitemShut {NoStop}%
\bibitem [{\citenamefont {Vl{\v{c}}ek}\ \emph {et~al.}(2016)\citenamefont
  {Vl{\v{c}}ek}, \citenamefont {Eisenberg}, \citenamefont {Steinle-Neumann},
  \citenamefont {Rabani}, \citenamefont {Neuhuaser},\ and\ \citenamefont
  {Baer}}]{Vlcek2016}%
  \BibitemOpen
  \bibfield  {author} {\bibinfo {author} {\bibfnamefont {V.}~\bibnamefont
  {Vl{\v{c}}ek}}, \bibinfo {author} {\bibfnamefont {H.~R.}\ \bibnamefont
  {Eisenberg}}, \bibinfo {author} {\bibfnamefont {G.}~\bibnamefont
  {Steinle-Neumann}}, \bibinfo {author} {\bibfnamefont {D.}~\bibnamefont
  {Neuhuaser}}, \bibinfo {author} {\bibfnamefont {E.}~\bibnamefont {Rabani}}, \
  and\ \bibinfo {author} {\bibfnamefont {R.}~\bibnamefont {Baer}},\ }\href@noop
  {} {\bibfield  {journal} {\bibinfo  {journal} {Phys. Rev. Lett.}\ }\textbf
  {\bibinfo {volume} {116}},\ \bibinfo {pages} {186401} (\bibinfo {year}
  {2016})}\BibitemShut {NoStop}%
\bibitem [{\citenamefont {Vl{\v{c}}ek}\ \emph {et~al.}(2018)\citenamefont
  {Vl{\v{c}}ek}, \citenamefont {Rabani},\ and\ \citenamefont
  {Neuhauser}}]{vlcek2018quasiparticle}%
  \BibitemOpen
  \bibfield  {author} {\bibinfo {author} {\bibfnamefont {V.}~\bibnamefont
  {Vl{\v{c}}ek}}, \bibinfo {author} {\bibfnamefont {E.}~\bibnamefont {Rabani}},
  \ and\ \bibinfo {author} {\bibfnamefont {D.}~\bibnamefont {Neuhauser}},\
  }\href@noop {} {\bibfield  {journal} {\bibinfo  {journal} {Phys. Rev. Matter.}\
  }\textbf {\bibinfo {volume} {2}},\ \bibinfo {pages} {030801} (\bibinfo {year}
  {2018})}\BibitemShut {NoStop}%
\bibitem [{\citenamefont {Filip}\ and\ \citenamefont
  {Giustino}(2014)}]{Filip2014}%
  \BibitemOpen
  \bibfield  {author} {\bibinfo {author} {\bibfnamefont {M.~R.}\ \bibnamefont
  {Filip}}\ and\ \bibinfo {author} {\bibfnamefont {F.}~\bibnamefont
  {Giustino}},\ }\href {\doibase 10.1103/PhysRevB.90.245145} {\bibfield
  {journal} {\bibinfo  {journal} {Phys. Rev. B}\ }\textbf {\bibinfo {volume}
  {90}},\ \bibinfo {pages} {245145} (\bibinfo {year} {2014})}\BibitemShut
  {NoStop}%
\bibitem [{\citenamefont {Qian}\ \emph {et~al.}(2015)\citenamefont {Qian},
  \citenamefont {Umari},\ and\ \citenamefont {Marzari}}]{Qian2015}%
  \BibitemOpen
  \bibfield  {author} {\bibinfo {author} {\bibfnamefont {X.}~\bibnamefont
  {Qian}}, \bibinfo {author} {\bibfnamefont {P.}~\bibnamefont {Umari}}, \ and\
  \bibinfo {author} {\bibfnamefont {N.}~\bibnamefont {Marzari}},\ }\href
  {\doibase 10.1103/PhysRevB.91.245105} {\bibfield  {journal} {\bibinfo
  {journal} {Phys. Rev. B}\ }\textbf {\bibinfo {volume} {91}},\ \bibinfo
  {pages} {245105} (\bibinfo {year} {2015})}\BibitemShut {NoStop}%
\bibitem [{\citenamefont {Martyna}\ and\ \citenamefont
  {Tuckerman}(1999)}]{martyna1999reciprocal}%
  \BibitemOpen
  \bibfield  {author} {\bibinfo {author} {\bibfnamefont {G.~J.}\ \bibnamefont
  {Martyna}}\ and\ \bibinfo {author} {\bibfnamefont {M.~E.}\ \bibnamefont
  {Tuckerman}},\ }\href@noop {} {\bibfield  {journal} {\bibinfo  {journal} {J. Chem. Phys.}\ }\textbf {\bibinfo {volume} {110}},\ \bibinfo
  {pages} {2810} (\bibinfo {year} {1999})}\BibitemShut {NoStop}%
\bibitem [{\citenamefont {Perdew}\ and\ \citenamefont
  {Wang}(1992)}]{PerdewWang}%
  \BibitemOpen
  \bibfield  {author} {\bibinfo {author} {\bibfnamefont {J.~P.}\ \bibnamefont
  {Perdew}}\ and\ \bibinfo {author} {\bibfnamefont {Y.}~\bibnamefont {Wang}},\
  }\href {\doibase 10.1103/PhysRevB.45.13244} {\bibfield  {journal} {\bibinfo
  {journal} {Phys. Rev. B}\ }\textbf {\bibinfo {volume} {45}},\ \bibinfo
  {pages} {13244} (\bibinfo {year} {1992})}\BibitemShut {NoStop}%
\bibitem [{\citenamefont {Troullier}\ and\ \citenamefont
  {Martins}(1991)}]{TroullierMartins1991}%
  \BibitemOpen
  \bibfield  {author} {\bibinfo {author} {\bibfnamefont {N.}~\bibnamefont
  {Troullier}}\ and\ \bibinfo {author} {\bibfnamefont {J.~L.}\ \bibnamefont
  {Martins}},\ }\href@noop {} {\bibfield  {journal} {\bibinfo  {journal} {Phys.
  Rev. B}\ }\textbf {\bibinfo {volume} {43}},\ \bibinfo {pages} {1993}
  (\bibinfo {year} {1991})}\BibitemShut {NoStop}%
\bibitem [{\citenamefont {Rangel}\ \emph {et~al.}(2016)\citenamefont {Rangel},
  \citenamefont {Hamed}, \citenamefont {Bruneval},\ and\ \citenamefont
  {Neaton}}]{rangel2016evaluating}%
  \BibitemOpen
  \bibfield  {author} {\bibinfo {author} {\bibfnamefont {T.}~\bibnamefont
  {Rangel}}, \bibinfo {author} {\bibfnamefont {S.~M.}\ \bibnamefont {Hamed}},
  \bibinfo {author} {\bibfnamefont {F.}~\bibnamefont {Bruneval}}, \ and\
  \bibinfo {author} {\bibfnamefont {J.~B.}\ \bibnamefont {Neaton}},\
  }\href@noop {} {\bibfield  {journal} {\bibinfo  {journal} {JJ. Chem. Theory Comput.}\ }\textbf {\bibinfo {volume} {12}},\
  \bibinfo {pages} {2834} (\bibinfo {year} {2016})}\BibitemShut {NoStop}%
\bibitem [{\citenamefont {van Setten}\ \emph
  {et~al.}(2015{\natexlab{b}})\citenamefont {van Setten}, \citenamefont
  {Caruso}, \citenamefont {Sharifzadeh}, \citenamefont {Ren}, \citenamefont
  {Scheffler}, \citenamefont {Liu}, \citenamefont {Lischner}, \citenamefont
  {Lin}, \citenamefont {Deslippe}, \citenamefont {Louie} \emph
  {et~al.}}]{vansetten2015gw}%
  \BibitemOpen
  \bibfield  {author} {\bibinfo {author} {\bibfnamefont {M.~J.}\ \bibnamefont
  {van Setten}}, \bibinfo {author} {\bibfnamefont {F.}~\bibnamefont {Caruso}},
  \bibinfo {author} {\bibfnamefont {S.}~\bibnamefont {Sharifzadeh}}, \bibinfo
  {author} {\bibfnamefont {X.}~\bibnamefont {Ren}}, \bibinfo {author}
  {\bibfnamefont {M.}~\bibnamefont {Scheffler}}, \bibinfo {author}
  {\bibfnamefont {F.}~\bibnamefont {Liu}}, \bibinfo {author} {\bibfnamefont
  {J.}~\bibnamefont {Lischner}}, \bibinfo {author} {\bibfnamefont
  {L.}~\bibnamefont {Lin}}, \bibinfo {author} {\bibfnamefont {J.~R.}\
  \bibnamefont {Deslippe}}, \bibinfo {author} {\bibfnamefont {S.~G.}\
  \bibnamefont {Louie}},  \emph {et~al.},\ }\href@noop {} {\bibfield  {journal}
  {\bibinfo  {journal} {J. Chem. Theory Comput.}\ }\textbf {\bibinfo {volume}
  {11}},\ \bibinfo {pages} {5665} (\bibinfo {year}
  {2015}{\natexlab{b}})}\BibitemShut {NoStop}%
\bibitem [{\citenamefont {Krause}\ \emph {et~al.}(2015)\citenamefont {Krause},
  \citenamefont {Harding},\ and\ \citenamefont {Klopper}}]{krause2015coupled}%
  \BibitemOpen
  \bibfield  {author} {\bibinfo {author} {\bibfnamefont {K.}~\bibnamefont
  {Krause}}, \bibinfo {author} {\bibfnamefont {M.~E.}\ \bibnamefont {Harding}},
  \ and\ \bibinfo {author} {\bibfnamefont {W.}~\bibnamefont {Klopper}},\
  }\href@noop {} {\bibfield  {journal} {\bibinfo  {journal} {Mol. Phys.}\
  }\textbf {\bibinfo {volume} {113}},\ \bibinfo {pages} {1952} (\bibinfo {year}
  {2015})}\BibitemShut {NoStop}%
\bibitem [{NIS()}]{NIST}%
  \BibitemOpen
  \href@noop {} {\enquote {\bibinfo {title} {Nist computational chemistry
  comparison and benchmark database nist standard reference database, number
  101; johnson, r. d., iii, ed.; 2016. http://cccbdb.nist.gov/},}\
  }\BibitemShut {NoStop}%
\bibitem [{\citenamefont {Vl\v{c}ek}\ \emph {et~al.}(2018)\citenamefont
  {Vl\v{c}ek}, \citenamefont {Rabani},\ and\ \citenamefont
  {Neuhauser}}]{Vlcek20182b}%
  \BibitemOpen
  \bibfield  {author} {\bibinfo {author} {\bibfnamefont {V.}~\bibnamefont
  {Vl\v{c}ek}}, \bibinfo {author} {\bibfnamefont {E.}~\bibnamefont {Rabani}}, \
  and\ \bibinfo {author} {\bibfnamefont {D.}~\bibnamefont {Neuhauser}},\
  }\href@noop {} {\bibfield  {journal} {\bibinfo  {journal} {to be submitted}\
  } (\bibinfo {year} {2018})}\BibitemShut {NoStop}%
\bibitem [{\citenamefont {van~Houcke}\ \emph {et~al.}(2017)\citenamefont
  {van~Houcke}, \citenamefont {Tupitsyn}, \citenamefont {Mishchenko},\ and\
  \citenamefont {Prokof'ev}}]{VanHoucke2017}%
  \BibitemOpen
  \bibfield  {author} {\bibinfo {author} {\bibfnamefont {K.}~\bibnamefont
  {van~Houcke}}, \bibinfo {author} {\bibfnamefont {I.~S.}\ \bibnamefont
  {Tupitsyn}}, \bibinfo {author} {\bibfnamefont {A.~S.}\ \bibnamefont
  {Mishchenko}}, \ and\ \bibinfo {author} {\bibfnamefont {N.~V.}\ \bibnamefont
  {Prokof'ev}},\ }\href {\doibase 10.1103/PhysRevB.95.195131} {\bibfield
  {journal} {\bibinfo  {journal} {Phys. Rev. B}\ }\textbf {\bibinfo {volume}
  {95}},\ \bibinfo {pages} {195131} (\bibinfo {year} {2017})}\BibitemShut
  {NoStop}%
\bibitem [{\citenamefont {Tiago}\ \emph {et~al.}(2004)\citenamefont {Tiago},
  \citenamefont {Ismail-Beigi},\ and\ \citenamefont {Louie}}]{tiago2004effect}%
  \BibitemOpen
  \bibfield  {author} {\bibinfo {author} {\bibfnamefont {M.~L.}\ \bibnamefont
  {Tiago}}, \bibinfo {author} {\bibfnamefont {S.}~\bibnamefont {Ismail-Beigi}},
  \ and\ \bibinfo {author} {\bibfnamefont {S.~G.}\ \bibnamefont {Louie}},\
  }\href@noop {} {\bibfield  {journal} {\bibinfo  {journal} {Phys. Rev. B}\
  }\textbf {\bibinfo {volume} {69}},\ \bibinfo {pages} {125212} (\bibinfo
  {year} {2004})}\BibitemShut {NoStop}%
\bibitem [{\citenamefont {Hybertsen}\ and\ \citenamefont
  {Louie}(1986{\natexlab{b}})}]{HybertsenLouie}%
  \BibitemOpen
  \bibfield  {author} {\bibinfo {author} {\bibfnamefont {M.~S.}\ \bibnamefont
  {Hybertsen}}\ and\ \bibinfo {author} {\bibfnamefont {S.~G.}\ \bibnamefont
  {Louie}},\ }\href@noop {} {\bibfield  {journal} {\bibinfo  {journal} {Phys.
  Rev. B}\ }\textbf {\bibinfo {volume} {34}},\ \bibinfo {pages} {5390}
  (\bibinfo {year} {1986}{\natexlab{b}})}\BibitemShut {NoStop}%
\bibitem [{\citenamefont {Godby}\ and\ \citenamefont
  {Needs}(1989)}]{godby1989metal}%
  \BibitemOpen
  \bibfield  {author} {\bibinfo {author} {\bibfnamefont {R.~W.}~\bibnamefont
  {Godby}}\ and\ \bibinfo {author} {\bibfnamefont {R.~J.}~\bibnamefont {Needs}},\
  }\href@noop {} {\bibfield  {journal} {\bibinfo  {journal} {Phys. Rev. Lett.}\ }\textbf {\bibinfo {volume} {62}},\ \bibinfo {pages} {1169}
  (\bibinfo {year} {1989})}\BibitemShut {NoStop}%
\bibitem [{\citenamefont {Larson}\ \emph {et~al.}(2013)\citenamefont {Larson},
  \citenamefont {Dvorak},\ and\ \citenamefont {Wu}}]{larson2013role}%
  \BibitemOpen
  \bibfield  {author} {\bibinfo {author} {\bibfnamefont {P.}~\bibnamefont
  {Larson}}, \bibinfo {author} {\bibfnamefont {M.}~\bibnamefont {Dvorak}}, \
  and\ \bibinfo {author} {\bibfnamefont {Z.}~\bibnamefont {Wu}},\ }\href@noop
  {} {\bibfield  {journal} {\bibinfo  {journal} {Phys. Rev. B}\ }\textbf
  {\bibinfo {volume} {88}},\ \bibinfo {pages} {125205} (\bibinfo {year}
  {2013})}\BibitemShut {NoStop}%
\bibitem [{\citenamefont {Bimberg}\ \emph {et~al.}(1981)\citenamefont
  {Bimberg}, \citenamefont {Altarelli},\ and\ \citenamefont
  {Lipari}}]{bimberg1981calculation}%
  \BibitemOpen
  \bibfield  {author} {\bibinfo {author} {\bibfnamefont {D.}~\bibnamefont
  {Bimberg}}, \bibinfo {author} {\bibfnamefont {M.}~\bibnamefont {Altarelli}},
  \ and\ \bibinfo {author} {\bibfnamefont {N.}~\bibnamefont {Lipari}},\
  }\href@noop {} {\bibfield  {journal} {\bibinfo  {journal} {Solid State
  Coimm.}\ }\textbf {\bibinfo {volume} {40}},\ \bibinfo {pages} {437}
  (\bibinfo {year} {1981})}\BibitemShut {NoStop}%
\bibitem [{\citenamefont {Towns}\ \emph {et~al.}(2014)\citenamefont {Towns},
  \citenamefont {Cockerill}, \citenamefont {Dahan}, \citenamefont {Foster},
  \citenamefont {Gaither}, \citenamefont {Grimshaw}, \citenamefont {Hazlewood},
  \citenamefont {Lathrop}, \citenamefont {Lifka}, \citenamefont {Peterson}
  \emph {et~al.}}]{towns2014xsede}%
  \BibitemOpen
  \bibfield  {author} {\bibinfo {author} {\bibfnamefont {J.}~\bibnamefont
  {Towns}}, \bibinfo {author} {\bibfnamefont {T.}~\bibnamefont {Cockerill}},
  \bibinfo {author} {\bibfnamefont {M.}~\bibnamefont {Dahan}}, \bibinfo
  {author} {\bibfnamefont {I.}~\bibnamefont {Foster}}, \bibinfo {author}
  {\bibfnamefont {K.}~\bibnamefont {Gaither}}, \bibinfo {author} {\bibfnamefont
  {A.}~\bibnamefont {Grimshaw}}, \bibinfo {author} {\bibfnamefont
  {V.}~\bibnamefont {Hazlewood}}, \bibinfo {author} {\bibfnamefont
  {S.}~\bibnamefont {Lathrop}}, \bibinfo {author} {\bibfnamefont
  {D.}~\bibnamefont {Lifka}}, \bibinfo {author} {\bibfnamefont {G.~D.}\
  \bibnamefont {Peterson}},  \emph {et~al.},\ }\href@noop {} {\bibfield
  {journal} {\bibinfo  {journal} {Computing in Science \& Engineering}\
  }\textbf {\bibinfo {volume} {16}},\ \bibinfo {pages} {62} (\bibinfo {year}
  {2014})}\BibitemShut {NoStop}%
\end{thebibliography}
\addcontentsline{toc}{section}{\refname}

\end{document}